\documentclass[11pt,a4paper]{article}
\pdfoutput=1
\usepackage[utf8]{inputenc}
\usepackage[english]{babel}

\usepackage{amsmath}
\usepackage{amsfonts}
\usepackage{amssymb}
\usepackage{graphicx}

\usepackage{caption}
\usepackage{subcaption}
\usepackage{float}
\usepackage{appendix}

\numberwithin{equation}{section}

\usepackage[hidelinks]{hyperref}

\newcommand\nn{\nonumber}
\newcommand\be{\begin{equation}}
\newcommand\ee{\end{equation}}
\newcommand\ba{\begin{eqnarray}}    %eqnarray
\newcommand\ea{\end{eqnarray}}      %eqnarray

\usepackage[left=2cm,right=2cm,top=2cm,bottom=2cm]{geometry}

\begin{document}
\begin{titlepage}

%%%%%%%%%%%%%%%%%%%%%%%%%%%%%%%%%%%%%%%%%%%%%%%%%%%%%%%%%%%%%%%%%%%%%%%%%%%%%%%%%%%%%%%%%%%%%%%%%%%%%%%%%%%%%%%%%%%%%%%%%%%%
%%%%%%%%%%%%%%%%%%%%%%%%%%%%%%%%%%%%%%%%%%%%%%%%%%%%%%%%%%%%%%%%%%%%%%%%%%%%%%%%%%%%%%%%%%%%%%%%%%%%%%%%%%%%%%%%%%%%%%%%%%%%%
%%%%%%%%%%%%%%%%%%%%%%%%%%%%%%%%%%%%%%%%%%%%%%%%%%%%%%%%%%%%%%%%%%%%%%%%%%%%%%%%%%%%%%%%%%%%%%%%%%%%%%%%%%%%%%%%%%%%%%%%%%%%%

\begin{center}
{\Large \bf   On excited states in real-time AdS/CFT}\\
\end{center}

%%%%%%%%%%%%%%%%%%%%%%%%%%%%%%%%%%%%%%%%%%%%%%%%%%%%%%%%%%%%%%%%%%%%%%%%%%%%%%%%%%%%%%%%%%%%%%%%%%%%%%%%%%%%%%%%%%%%%%%%%%%%
%%%%%%%%%%%%%%%%%%%%%%%%%%%%%%%%%%%%%%%%%%%%%%%%%%%%%%%%%%%%%%%%%%%%%%%%%%%%%%%%%%%%%%%%%%%%%%%%%%%%%%%%%%%%%%%%%%%%%%%%%%%%%
%%%%%%%%%%%%%%%%%%%%%%%%%%%%%%%%%%%%%%%%%%%%%%%%%%%%%%%%%%%%%%%%%%%%%%%%%%%%%%%%%%%%%%%%%%%%%%%%%%%%%%%%%%%%%%%%%%%%%%%%%%%%%

\vskip .8 cm
\begin{center}
%{\large
Marcelo Botta-Cantcheff , Pedro J. Mart\'inez and Guillermo A. Silva

\vskip .3 cm
{\it  Instituto de F\'\i sica de La Plata - CONICET \& \\
Departamento de F\'\i sica - UNLP}
{\it C.C. 67, 1900 La Plata, Argentina}
\vskip .3 cm

E-mail: botta@fisica.unlp.edu.ar, martinezp@fisica.unlp.edu.ar, silva@fisica.unlp.edu.ar
\end{center}

\vskip .8 cm
\begin{abstract}
The Skenderis-van Rees prescription, which allows the calculation of time-ordered correlation functions of local operators in CFT's using holographic methods is studied and applied for excited states. Calculation of correlators and matrix elements of local CFT operators between generic in/out
states are carried out in global Lorentzian AdS. We find the precise form of such states, obtain an holographic formula to compute the inner product between them, and using the consistency with other known  prescriptions, we argue that the in/out excited states built according to the Skenderis-Van Rees prescription correspond to {\it coherent} states in the (large-$N$) AdS-Hilbert space. This is confirmed by explicit holographic computations.
The outcome of this study has remarkable implications on generalizing the Hartle-Hawking construction for wave functionals of excited states in AdS quantum gravity.
\end{abstract}
%%%%%%%%%%%%%%%%%%%%%%%%%%%%%%%%%%%%%%%%%%%%%%%%%%%%%%%%%%%%%%%%%%%%%%%%%%%%%%%%%%%%%%%%%%%%%%%%%%%%%%%%%%%%%%%%%%%%%%%%%%%%
%%%%%%%%%%%%%%%%%%%%%%%%%%%%%%%%%%%%%%%%%%%%%%%%%%%%%%%%%%%%%%%%%%%%%%%%%%%%%%%%%%%%%%%%%%%%%%%%%%%%%%%%%%%%%%%%%%%%%%%%%%%%%
%%%%%%%%%%%%%%%%%%%%%%%%%%%%%%%%%%%%%%%%%%%%%%%%%%%%%%%%%%%%%%%%%%%%%%%%%%%%%%%%%%%%%%%%%%%%%%%%%%%%%%%%%%%%%%%%%%%%%%%%%%%%%

\tableofcontents

%%%%%%%%%%%%%%%%%%%%%%%%%%%%%%%%%%%%%%%%%%%%%%%%%%%%%%%%%%%%%%%%%%%%%%%%%%%%%%%%%%%%%%%%%%%%%%%%%%%%%%%%%%%%%%%%%%%%%%%%%%%%
%%%%%%%%%%%%%%%%%%%%%%%%%%%%%%%%%%%%%%%%%%%%%%%%%%%%%%%%%%%%%%%%%%%%%%%%%%%%%%%%%%%%%%%%%%%%%%%%%%%%%%%%%%%%%%%%%%%%%%%%%%%%%
%%%%%%%%%%%%%%%%%%%%%%%%%%%%%%%%%%%%%%%%%%%%%%%%%%%%%%%%%%%%%%%%%%%%%%%%%%%%%%%%%%%%%%%%%%%%%%%%%%%%%%%%%%%%%%%%%%%%%%%%%%%%%
\end{titlepage}
%%%%%%%%%%%%%%%%%%%%%%%%%%%%%%%%%%%%%%%%%%%%%%%%%%%%%%%%%%%%%%%%%%%%%%%%%%%%%%%%%%%%%%%%%%%%%%%%%%%%%%%%%%%%%%%%%%%%%%%%%%%%
%%%%%%%%%%%%%%%%%%%%%%%%%%%%%%%%%%%%%%%%%%%%%%%%%%%%%%%%%%%%%%%%%%%%%%%%%%%%%%%%%%%%%%%%%%%%%%%%%%%%%%%%%%%%%%%%%%%%%%%%%%%%%
%%%%%%%%%%%%%%%%%%%%%%%%%%%%%%%%%%%%%%%%%%%%%%%%%%%%%%%%%%%%%%%%%%%%%%%%%%%%%%%%%%%%%%%%%%%%%%%%%%%%%%%%%%%%%%%%%%%%%%%%%%%%%

\section{Introduction}

%%%%%%%%%%%%%%%%%%%%%%%%%%%%%%%%%%%%%%%%%%%%%%%%%%%%%%%%%%%%%%%%%%%%%%%%%%%%%%%%%%%%%%%%%%%%%%%%%%%%%%%%%%%%%%%%%%%%%%%%%%%%
%%%%%%%%%%%%%%%%%%%%%%%%%%%%%%%%%%%%%%%%%%%%%%%%%%%%%%%%%%%%%%%%%%%%%%%%%%%%%%%%%%%%%%%%%%%%%%%%%%%%%%%%%%%%%%%%%%%%%%%%%%%%%
%%%%%%%%%%%%%%%%%%%%%%%%%%%%%%%%%%%%%%%%%%%%%%%%%%%%%%%%%%%%%%%%%%%%%%%%%%%%%%%%%%%%%%%%%%%%%%%%%%%%%%%%%%%%%%%%%%%%%%%%%%%%%

\begin{figure}[t]\centering
\begin{subfigure}{0.49\textwidth}\centering
\includegraphics[width=.9\linewidth] {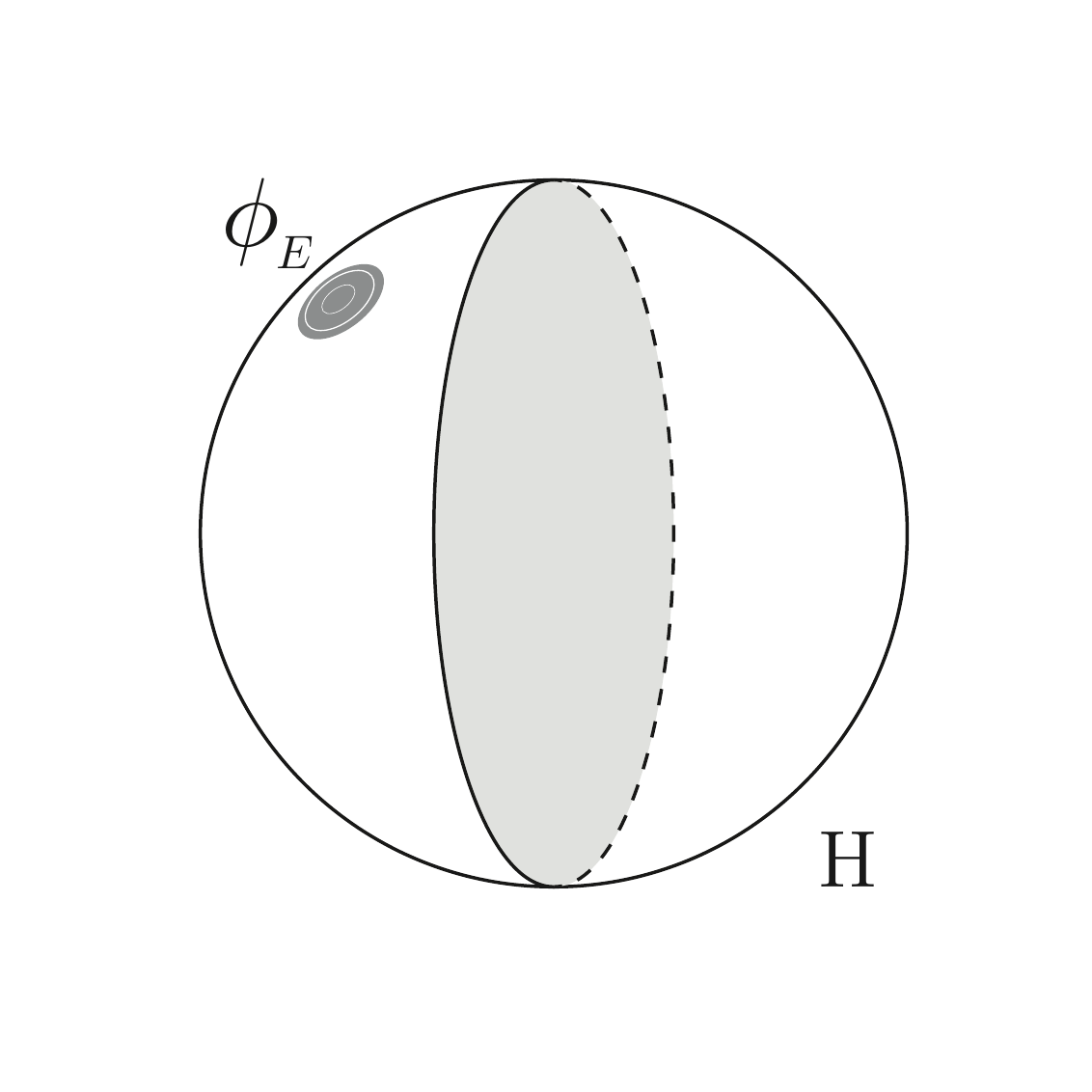}
\caption{}
\end{subfigure}
\begin{subfigure}{0.49\textwidth}\centering
\includegraphics[width=.9\linewidth] {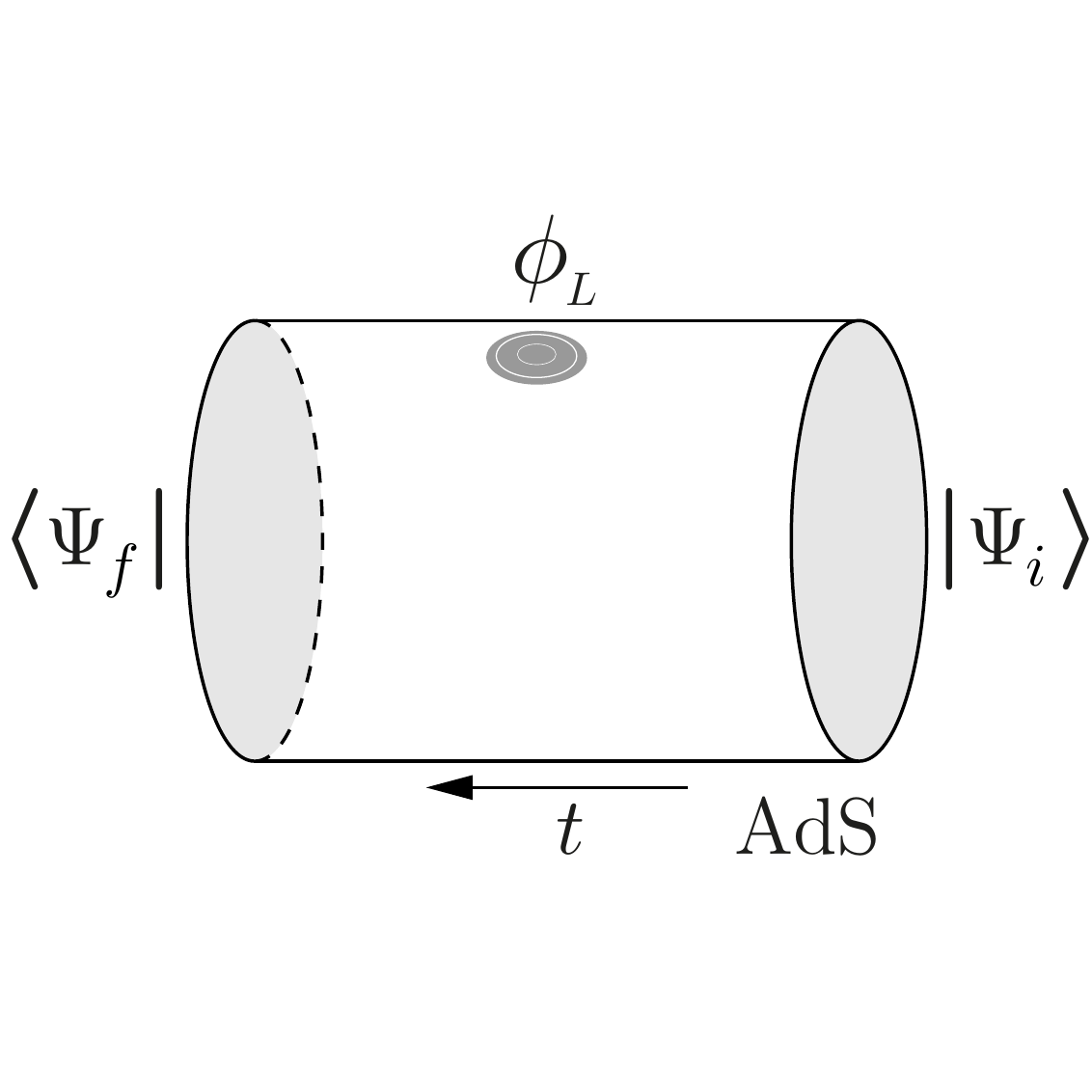}
\caption{}
\end{subfigure}
\caption{(a) Euclidean AdS (hyperbolic space) and the insertion of a source at its conformal boundary.
(b) Lorentzian AdS with a source at its boundary. We also depict the initial and final wave functions. }
\label{Sources}
\end{figure}

The Skenderis and van Rees (SvR) prescription \cite{SvRC,SvRL} allows the calculation in real time of $n$-point
correlation functions of local CFT operator $\mathcal{O}$ using the gauge/gravity correspondence. It can be
understood as an extension of the GKPW prescription  which in the supergravity approximation can be summarized as \cite{GKP1,GKP2}
\begin{equation}
 \langle\, e^{-\int_{\partial H}  \mathcal{O} \,\phi_E } \,\rangle \equiv e^{-S^{0}_E[\phi_E]}\,.
 \label{GKP}
\end{equation}
The integration over $\partial$H in the lhs, giving the generating functional for $\cal O$ correlation functions,
suggests  the well known statement that the Euclidean CFT lives in the
boundary of the bulk H (hyperbolic or Euclidean AdS) space shown in figure  \ref{Sources}a.  $S_E^0[\phi_E]$ on the rhs is the on-shell action
for the bulk field $\Phi$, dual to  $\cal O$, having $\phi_E$ as boundary condition on the conformal boundary $\partial$H.
The bulk field boundary condition $\phi_E$ acts as the source for the dual CFT operator $\cal O$.

\begin{figure}[t]\centering
\begin{subfigure}{0.49\textwidth}\centering
\includegraphics[width=.9\linewidth] {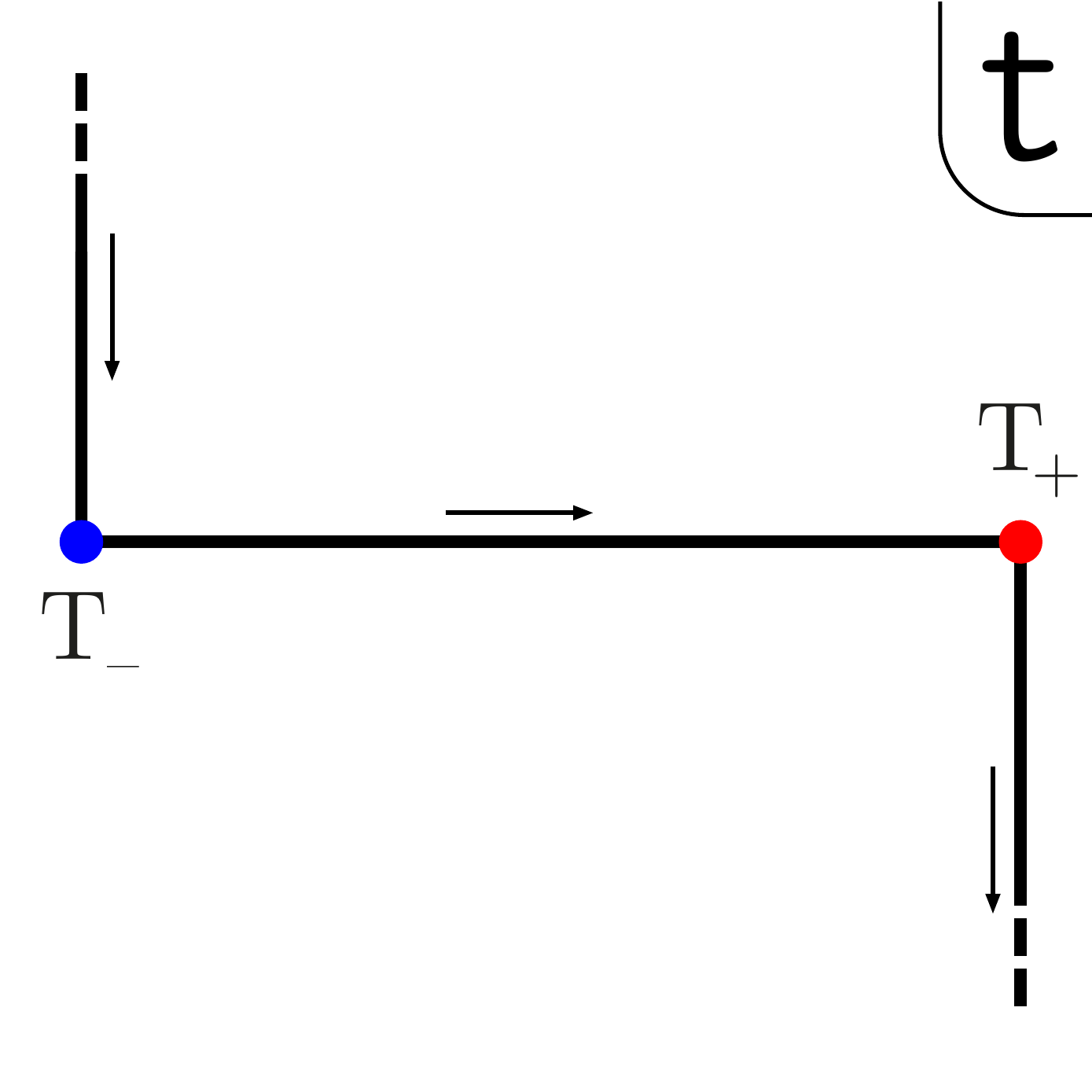}
\caption{}
\end{subfigure}
\begin{subfigure}{0.49\textwidth}\centering
\includegraphics[width=.9\linewidth] {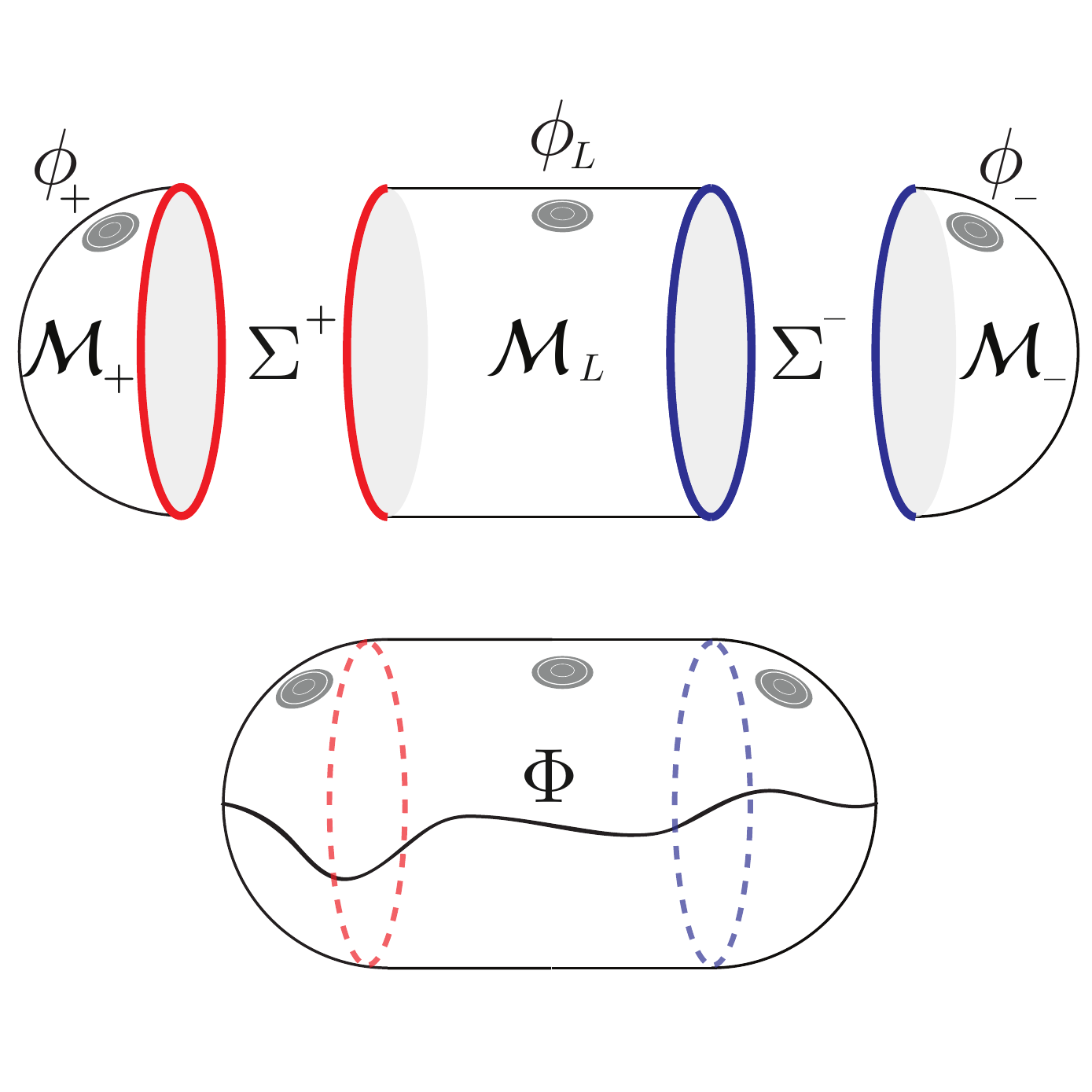}
\caption{}
\end{subfigure}
\caption{(a) In-Out contour in complex t-plane appropriate for scattering problems showing a temporal evolution $\Delta T=T_+-T_-$. (b) SvR geometry dual to In-Out QFT contour
obtained by gluing together Lorentzian ${\cal M}_L$ and Euclidean ${\cal M_\pm}$ AdS sections. We explicitly
depict the gluing surfaces $\Sigma^\pm$. On the bottom we show the glued geometry and a generic smooth configuration on the spacetime.}
\label{SvR}
\end{figure}

A  generalization of \eqref{GKP} to real time situations requires the specification of both the initial and final states in
addition to the boundary condition $\phi_L$ on the timelike boundary (see figure  \ref{Sources}b)\cite{ABS}.
In their original work SvR gave a prescription for building out the CFT vacuum state, both as initial and final states, in terms of
boundary conditions for the bulk fields on a manifold constructed out by gluing Lorentzian and Euclidean sections (see figure \ref{SvR}b).
From a QFT perspective one could envisage, at zero temperature, computing either scattering amplitudes or expectation values.
The appropriate formalism for these problems goes under the name of Schwinger-Keldysh and involve contours in the complex
$\sf t$-plane known as In-Out or In-In respectively\cite{KamRam}.  For scattering problems, the appropriate contour is
shown in figure \ref{SvR}(a), the prescription given by SvR yields

%\begin{figure}[t]\centering
%\begin{subfigure}{0.49\textwidth}\centering
%\includegraphics[width=.9\linewidth] {HBorde.pdf}
%\caption{}
%\end{subfigure}
%\begin{subfigure}{0.49\textwidth}\centering
%\includegraphics[width=.9\linewidth] {AdSBorde.pdf}
%\caption{}
%\end{subfigure}
%\caption{(a) Euclidean AdS (hyperbolic space) and the insertion of a source at its conformal boundary.
%(b) Lorentzian AdS with a source at its boundary. We also depict the initial and final wave functions. }
%\label{Sources}
%\end{figure}
%
%\begin{figure}[t]\centering
%\begin{subfigure}{0.49\textwidth}\centering
%\includegraphics[width=.9\linewidth] {EjemploAF.pdf}
%\caption{}
%\end{subfigure}
%\begin{subfigure}{0.49\textwidth}\centering
%\includegraphics[width=.9\linewidth] {EjemploCF.pdf}
%\caption{}
%\end{subfigure}
%\caption{(a) In-Out contour in complex t-plane appropriate for scattering problems showing a temporal evolution $\Delta T=T_+-T_-$. (b) SvR geometry dual to In-Out QFT contour
%obtained by gluing together Lorentzian ${\cal M}_L$ and Euclidean ${\cal M_\pm}$ AdS sections. We explicitly
%depict the gluing surfaces $\Sigma^\pm$. On the bottom we show the glued geometry and a generic smooth configuration on the spacetime.}
%\label{SvR}
%\end{figure}

\begin{equation}
\label{SvR:Main:SvR-Def}
\langle 0 | \, e^{-i \int_{\partial_r \mathcal{M}_L} \mathcal{O} \phi_L }\,  | 0 \rangle
\equiv e^{i S^0_L[\phi_L;\phi_{\Sigma^-},\phi_{\Sigma^+}] - S^0_-[0;\phi_{\Sigma^-}] - S^0_+[0;\phi_{\Sigma^+}]}\,.
\end{equation}
The lhs gives the generating function of time ordered correlation functions of $\cal O$ in a Lorentzian CFT that lives in the
timelike conformal boundary $\partial_r \mathcal{M}_L$ of the bulk spacetime. In the rhs, $S^0_L[\phi_L;\phi_{\Sigma^-},\phi_{\Sigma^+}]$
is the Lorentzian on-shell action for a bulk field $\Phi_L$ which takes boundary values $\phi_{\Sigma^\pm}$ on the spacelike boundaries
$\Sigma^{\pm}\equiv\partial_t{\cal M}_L$ and  $\phi_L$ over $\partial_r \mathcal{M}_{L}$. The exponents $S^0_{\pm}[0;\phi_{\Sigma^{\pm}}]$
are the bulk field on shell actions on the Euclidean sections $\mathcal{M}_{\pm}$ for boundary values $\phi_\pm=0$ on $\partial_r \mathcal{M}_{\pm}$
and $\phi_{\Sigma^\pm}$ on $\Sigma^{\pm}$ .

It is worth noticing that (\ref{SvR:Main:SvR-Def}) implicitly assumes the bulk fields, and its conjugated momenta,
to be continuous through the $\Sigma^\pm$ gluing surfaces. These conditions follows from a complete quantum treatment:
continuity of the fields is implicit in every path integral treatment, which also requires the integration of the rhs of
(\ref{SvR:Main:SvR-Def}) over all possible configurations $\phi_{\Sigma^{\pm}}$. In a semi-classical
approximation, the  leading  contribution arising from minimizing the (complex) on shell action with respect to the
boundary conditions $\phi_{\Sigma^\pm}$ leads to continuity of momenta.

In their original proposal, Skenderis and van-Rees have stressed that their prescription was in line with the Hartle-Hawking construction
\cite{HH}, which as we will review below, is implicitly assumed and plays a crucial role in prescribing the initial/final wave functionals
for the ground state of the gravity side.  Thus, one of the motivations for the present work is to learn from the AdS/CFT setup how
% So one of our main motivations with this study is to learn from the AdS/CFT laboratory, how
this construction is generalized to compute wave functions of excited states in (quantum) gravity.

The authors suggested in \cite{SvRC} that in order to describe non-vacuum states, one should consider turning on boundary conditions over $\partial_r \mathcal{M}_{\pm}$, but they didn't explore this idea any further.
In this paper, we pursue a  two-fold purpose: on the one hand we verify explicitly this
claim, and on a second hand we characterize the  properties of the (AdS/)CFT  states constructed by this prescription.
Non trivial boundary conditions $\phi_\pm\neq0$ will be turned on in the conformal boundaries $\partial_r{\cal M}_\pm$. As a result
we will confirm the SvR claim and,  in addition, the computation will show that the
resulting excitations correspond to {\it coherent} states. For previous work regarding excited states in Lorentzian signature
see \cite{avis,breitfreed,bala1,BDHM,BDHM2,bala2,bala3,satoh,marolf,sever,raulo}.

This article is organized as follows. In Sec. \ref{construct} %we briefly review the SvR method with an example for a free scalar field and explain the geometric construction related to the In-Out formalism. In Sec ... we revise
the SvR proposal is revised, {% and the arguments which it is based on are explained in detail.}
while in Sec. \ref{excited} the SvR claim is generalized for excited states and, arguing consistency with other well known
prescriptions existing in the literature \cite{BDHM}, it is argued that the initial and final states can be considered
\emph{coherent} in a AdS Fock space representation. In Sec. \ref{ExpVal} non vanishing boundary conditions  are imposed in the
Euclidean regions of the spacetime shown in figure  \ref{SvR}, and the main calculations carried out. Finally, in Sec \ref{check}
we check that our final results are consistent with our claim on coherent states. Concluding remarks are collected in Sec
\ref{concluding}, where we stress what this result learns us on generalizing the HH construction of quantum gravity states in arbitrary spacetime asymptotics.

%%%%%%%%%%%%%%%%%%%%%%%%%%%%%%%%%%%%%%%%%%%%%%%%%%%%%%%%%%%%%%%%%%%%%%%%%%%%%%%%%%%%%%%%%%%%%%%%%%%%%%%%%%%%%%%%%%%%%%%%%%%%
%%%%%%%%%%%%%%%%%%%%%%%%%%%%%%%%%%%%%%%%%%%%%%%%%%%%%%%%%%%%%%%%%%%%%%%%%%%%%%%%%%%%%%%%%%%%%%%%%%%%%%%%%%%%%%%%%%%%%%%%%%%%%
%%%%%%%%%%%%%%%%%%%%%%%%%%%%%%%%%%%%%%%%%%%%%%%%%%%%%%%%%%%%%%%%%%%%%%%%%%%%%%%%%%%%%%%%%%%%%%%%%%%%%%%%%%%%%%%%%%%%%%%%%%%%%

\section{Review of the SvR construction}
\label{construct}

%%%%%%%%%%%%%%%%%%%%%%%%%%%%%%%%%%%%%%%%%%%%%%%%%%%%%%%%%%%%%%%%%%%%%%%%%%%%%%%%%%%%%%%%%%%%%%%%%%%%%%%%%%%%%%%%%%%%%%%%%%%%
%%%%%%%%%%%%%%%%%%%%%%%%%%%%%%%%%%%%%%%%%%%%%%%%%%%%%%%%%%%%%%%%%%%%%%%%%%%%%%%%%%%%%%%%%%%%%%%%%%%%%%%%%%%%%%%%%%%%%%%%%%%%%
%%%%%%%%%%%%%%%%%%%%%%%%%%%%%%%%%%%%%%%%%%%%%%%%%%%%%%%%%%%%%%%%%%%%%%%%%%%%%%%%%%%%%%%%%%%%%%%%%%%%%%%%%%%%%%%%%%%%%%%%%%%%%

For a free field $\Phi$ on a (asymptotically) AdS Euclidean spacetime $\mathcal{M}_E$, the GKPW prescription \cite{GKP1,GKP2} beyond the semi-classical gravity approximation reads

\be\label{valoresperadoEUCL}
\langle \, e^{-\int_{\partial \mathcal{M}_E}  \mathcal{O}\, \phi_E }\,  \rangle \equiv \langle 0|  e^{-\int_{\partial \mathcal{M}_E}  \mathcal{O} \phi_E } |0 \rangle
= {\cal Z}[\phi_E]\equiv \int [\mathcal{D}\Phi]_{\phi_E} e^{-S_E[\Phi]}\,,
\ee
where   $[\mathcal{D}\Phi]_{\phi_E} $ denotes that the functional integral should be computed over configurations which satisfy $\Phi=\phi_E$ on the conformal boundary of ${\cal M}_E$.
Thus, the natural generalization of this formula to Lorentzian AdS spacetime involves vacuum wave functionals $\Psi_0$ as
\be\label{valoresperadoLorentz}
 \langle 0|\, T[ e^{-i\int_{\partial_r \mathcal{M}_L} \mathcal{O} \phi_L } ]\,|0 \rangle = \sum_{\phi_{\Sigma^\pm}} \,\left(\Psi_0[\phi_{\Sigma^+}]\,\right)^*\,{\cal Z}[\phi_L;\phi_{\Sigma^-}, \phi_{\Sigma^+}]\, \Psi_0[\phi_{\Sigma^-}]\,.
 \ee
The partition function in \eqref{valoresperadoEUCL} turns into the Feynman's path integral for the transition amplitude from an initial condition $\phi_{\Sigma^-}$ to a final condition $\phi_{\Sigma^+}$  at (spacelike) surfaces $ \Sigma^-$ and $ \Sigma^+$ respectively of the Lorentzian
AdS cylinder ${\cal M}_L$ shown in  figure  \ref{SvR}b,. In the Lorentzian setup $\phi_L$ denotes the (asymptotic) boundary
 condition for $\Phi$ at the timelike boundary
\be
\label{pathI}
{\cal Z}[\phi_L;\phi_{\Sigma^-}, \phi_{\Sigma^+}] \equiv \int_{\phi_{\Sigma^-}}^{\phi_{\Sigma^+}}[\mathcal{D}\Phi]_{\phi_L} \,\, e^{iS_L[\Phi]}.
\ee
The problem with this formula is that \emph{a priori}, one does not have a precise prescription for the values of $\phi_{\Sigma^\pm}$, nor the
form of the initial/final states described by the wave functionals  $\Psi^0[\phi_{\Sigma^\pm}]$ on the $\Sigma^\pm$ surfaces.
 Thus, the crucial steps of the SvR construction are:
firstly to identify the CFT vacuum $|0 \rangle$ with the wave functional of the fundamental state in the bulk theory
\be\nn
|0 \rangle \Leftrightarrow \Psi^0[\phi_{\Sigma^-}]\equiv\langle \, \phi_{\Sigma^{-}}\,|\,\Psi^0\, \rangle\,,
\ee
where it is implicitly assumed that $\phi_{\Sigma^\pm}$ on the spacelike surfaces $\Sigma^\pm$ constitutes a (configuration) basis for the Hilbert space of
the second-quantized bulk scalar field, so that the identity can be expressed as\footnote{More precisely, $|\phi_{\Sigma^\pm}\rangle$ are eigenstates of the field operator $\Phi$.}
\begin{equation}\label{identidad}
\sum_{\phi_{\Sigma^\pm}}\, |\,\phi_{\Sigma^\pm}\, \rangle \langle\,\phi_{\Sigma^\pm}\, |= 1\, .
\end{equation}
The rhs of (\ref{valoresperadoLorentz}) can therefore be rewritten as
\be\nn
\sum_{\phi_{\Sigma^\pm}} \,\left(\Psi^0[\phi_{\Sigma^+}]\,\right)^*
{\cal Z}[\phi_L;\phi_{\Sigma^-}, \phi_{\Sigma^+}]\,\, \Psi^0[\phi_{\Sigma^-}] =
\langle \Psi^0\,| \,{\cal U}[ T_- , T_+]_{\phi_L}\, |\,\Psi^0\rangle\,,
\ee
where ${\cal U}$ is the real-time evolution operator of the bulk theory.
Secondly, they use the Hartle-Hawking (HH) prescription for vacuum wave functionals \cite{HH}
\begin{equation}
\label{MGP:Pre:HH+}
\Psi^0[\phi_{\Sigma^{-}}] \equiv \int^{\phi_{\Sigma^{-}}}_{0} [\mathcal{D}\Phi]_{0} \, e^{-S_-[\Phi]}\,.
\end{equation}
This functional integral is computed summing over the field configurations on a half of
Euclidean AdS: $\mathcal{M}_{-}$ (see figure 2b and figure \ref{M+}).

Skenderis and Van Rees have defined this Euclidean path integral with timelike boundaries, like AdS, considering decaying field configurations at the asymptotic (radial) conformal boundary, and so the vacuum state is \emph{prepared} at the asymptotic boundary by imposing
vanishing boundary conditions. This is denoted by the $[\mathcal{D}\Phi^{}]_0$, where
  the subindex 0 instructs to sum over field configurations
with vanishing boundary conditions on $\mathcal{S}_{-}\equiv \partial_r\mathcal{M}_{-} \,- \,\{\tau = - \infty\}$, and the $0$ in the lower limit of the (path) integral
denotes the value of the field configuration at (Euclidean) infinite past  $\Phi (\tau = - \infty) $
  (see figure  \ref{M+}).
Moreover, SvR have also claimed that excited wave functionals \emph{would be}
obtained by imposing non-vanishing boundary conditions on the asymptotic  Euclidean boundary \cite{SvRC}. Their suggestion for an
excited wave functional can be expressed as
\begin{equation}
\label{wavef-g}
\Psi^{\phi_{-}}[\phi_{\Sigma^-}] \equiv \int^{\phi_{\Sigma^-}}_0 [\mathcal{D}\Phi]_{\phi_{-}} \,  e^{-S^{}_-[\Phi]}\,\, .
\end{equation}
Here the non-trivial smooth function $\phi_{-}$ stands for the value of $\Phi$ at $\mathcal{S}_{-}$,
and in what follows the value of $\Phi$ at the poles $\tau=\pm \infty$ is set to zero, motivated by the fact that in ordinary QFT this value only affects the normalization of the wave function.

\begin{figure}[t]\centering
\begin{subfigure}{0.49\textwidth}\centering
\includegraphics[width=.9\linewidth] {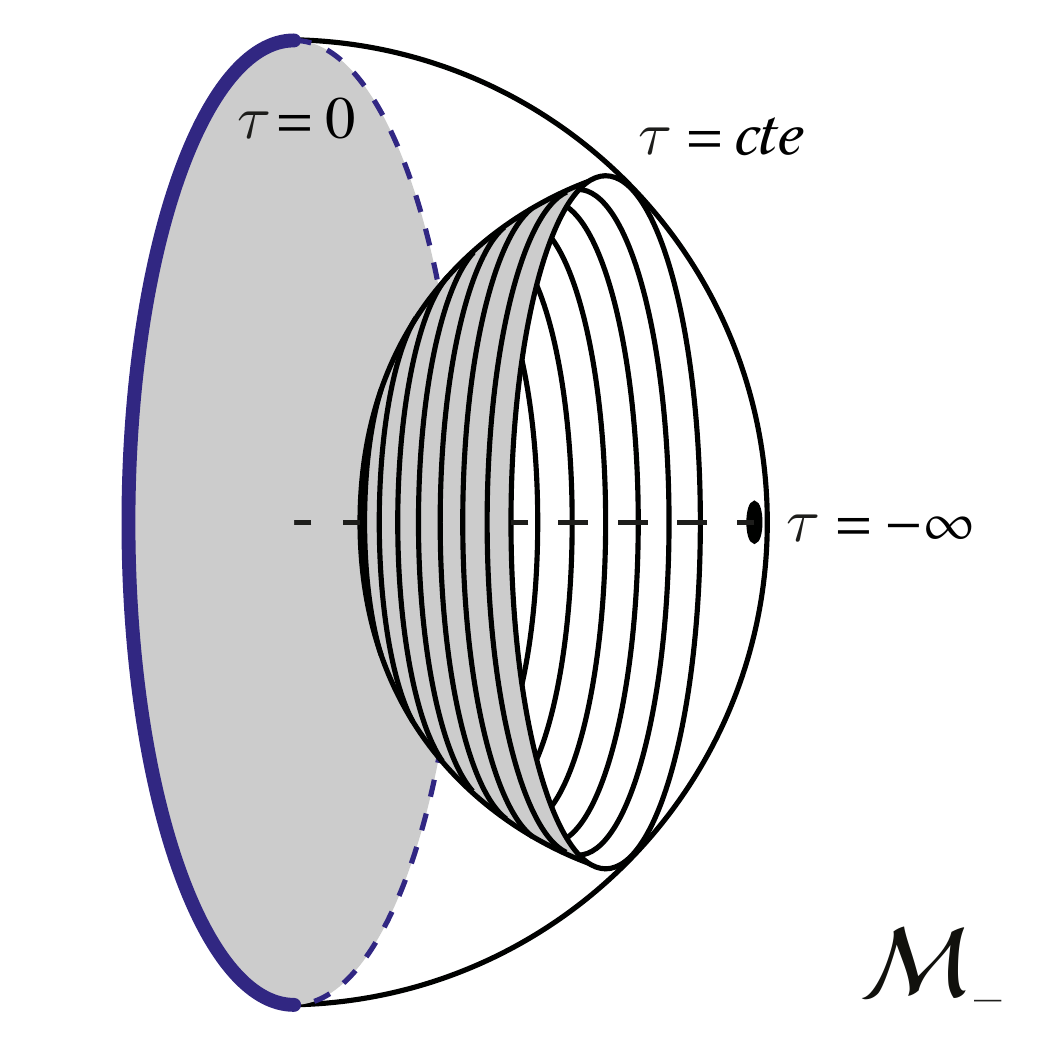}
\caption{}
\end{subfigure}
\begin{subfigure}{0.49\textwidth}\centering
\includegraphics[width=.9\linewidth] {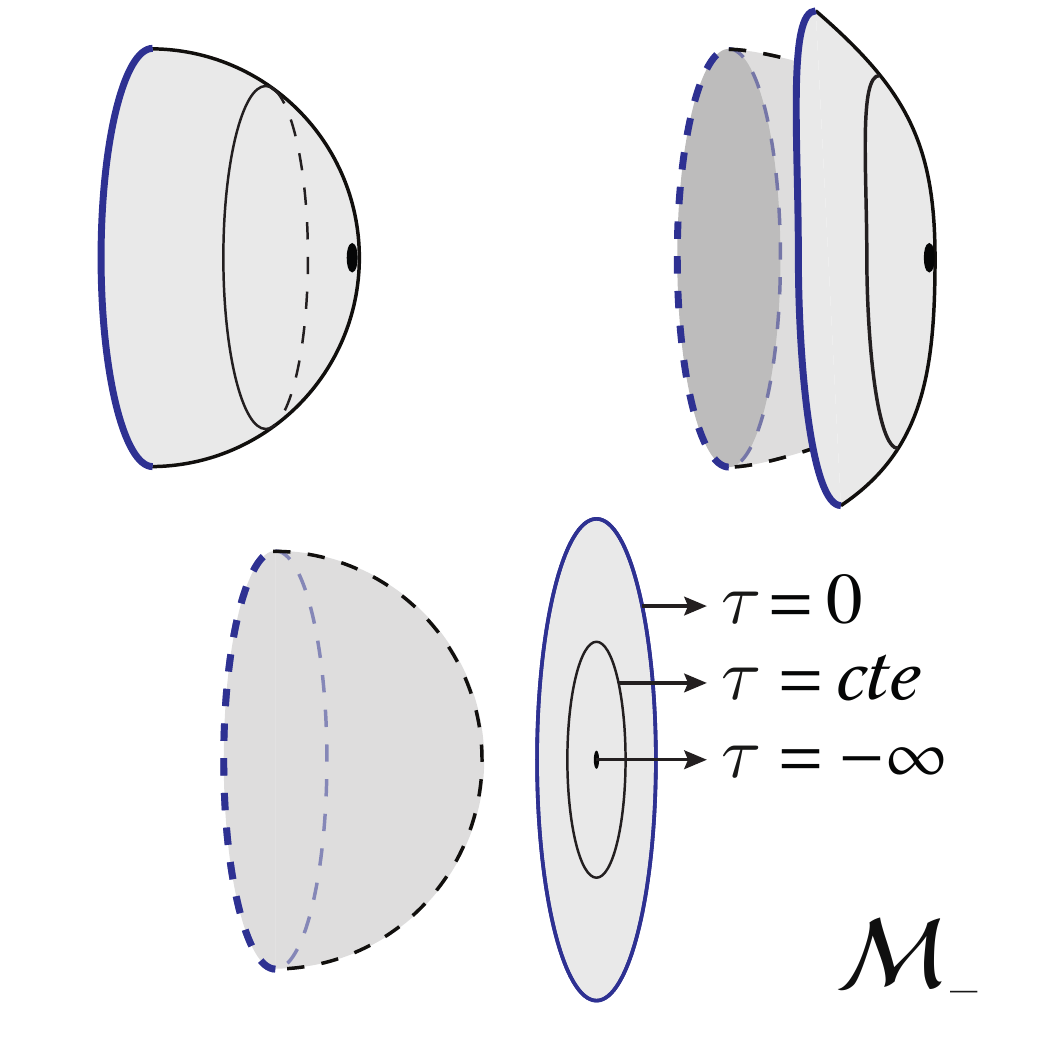}
\caption{}
\end{subfigure}
\caption{In standard global coordinates, the point $\tau = \pm \infty$ is mapped to the origin of the conformal boundary of $\mathcal{M}_\pm$.}
\label{M+}
\end{figure}%(see figure  \ref{MGP:Fig:M+}) %(\textbf{see Apendix})

Finally, plugging expression (\ref{pathI}) and (\ref{MGP:Pre:HH+}) into (\ref{valoresperadoLorentz}), results in
\be\label{SvR-path}
 \langle 0| T[ e^{-i\int_{\partial_r \mathcal{M}_L}  \mathcal{O} \,\phi_L} ]|0 \rangle = \sum_{\phi_{\Sigma^{\pm}}}
 \left(\int_{\phi_{\Sigma^{+}}}^{0} [\mathcal{D}\Phi]_{0}\, e^{-S_+[\Phi]}\right)
 \left(\int_{\phi_{\Sigma^{-}}}^{\phi_{\Sigma^{+}}} [\mathcal{D}\Phi]_{\phi_L} e^{ i S_L[\Phi]}\right)
  \left(\int_{0}^{\phi_{\Sigma^{-}}} [\mathcal{D}\Phi]_{0}\, e^{- S_-[\Phi]}\right)\,.
 \ee
Notice that the integration is taken over three intervals with different signatures, and therefore, the full action turns out to be
a complex-valued functional of the fields, and the r.h.s. of (\ref{SvR-path}) can be written as \emph{single} path integral over field configurations on an
AdS manifold with both Lorentzian and Euclidean pieces as shown in figure  \ref{SvR} b.

Performing a saddle-point approximation for each interval, one obtains
\be
 \langle 0| T[ e^{-i\int_{\partial_r \mathcal{M}_L} \mathcal{O}\, \phi_L } ]|0 \rangle = \sum_{\phi_{\Sigma^{\pm}}}\, e^{- S^0_-[0;\phi_{\Sigma^{-}}] + i S^0_L[\phi_L;\phi_{\Sigma^{-}},\phi_{\Sigma^{+}}] - S^0_{+}[0;\phi_{\Sigma^{+}}]}\,.
\ee%\label{SvR-path-saddle1}
The SvR prescription (\ref{SvR:Main:SvR-Def}) is recovered upon performing a second saddle point approximation which consists in finding the
term with maximal contribution to the $\phi_{\Sigma^{\pm}}$ sum. Minimizing the complex action w.r.t. the field values on the gluing surfaces, results in
\be
\label{glue}
\frac{\delta \,( -S^0_\pm + i S^0_L ) }{\delta \phi_{\Sigma^{\pm}}} = \pi_\pm + i \pi_L = 0\,.
\ee
For a free scalar field, this equation sets the continuity of the \emph{normal derivative} of the field through the gluing surfaces
$\Sigma^\pm$ \cite{SvRL}, which completes the SvR prescription.
The SvR proposal (\ref{wavef-g}) for excited states is precisely the statement that we are going to investigate in the forthcoming sections.

%%%%%%%%%%%%%%%%%%%%%%%%%%%%%%%%%%%%%%%%%%%%%%%%%%%%%%%%%%%%%%%%%%%%%%%%%%%%%%%%%%%%%%%%%%%%%%%%%%%%%%%%%%%%%%%%%%%%%%%%%%%%
%%%%%%%%%%%%%%%%%%%%%%%%%%%%%%%%%%%%%%%%%%%%%%%%%%%%%%%%%%%%%%%%%%%%%%%%%%%%%%%%%%%%%%%%%%%%%%%%%%%%%%%%%%%%%%%%%%%%%%%%%%%%%
%%%%%%%%%%%%%%%%%%%%%%%%%%%%%%%%%%%%%%%%%%%%%%%%%%%%%%%%%%%%%%%%%%%%%%%%%%%%%%%%%%%%%%%%%%%%%%%%%%%%%%%%%%%%%%%%%%%%%%%%%%%%%
\section{Excited $\Psi^{\phi_\pm}$ states}
\label{excited}
%%%%%%%%%%%%%%%%%%%%%%%%%%%%%%%%%%%%%%%%%%%%%%%%%%%%%%%%%%%%%%%%%%%%%%%%%%%%%%%%%%%%%%%%%%%%%%%%%%%%%%%%%%%%%%%%%%%%%%%%%%%%
%%%%%%%%%%%%%%%%%%%%%%%%%%%%%%%%%%%%%%%%%%%%%%%%%%%%%%%%%%%%%%%%%%%%%%%%%%%%%%%%%%%%%%%%%%%%%%%%%%%%%%%%%%%%%%%%%%%%%%%%%%%%%
%%%%%%%%%%%%%%%%%%%%%%%%%%%%%%%%%%%%%%%%%%%%%%%%%%%%%%%%%%%%%%%%%%%%%%%%%%%%%%%%%%%%%%%%%%%%%%%%%%%%%%%%%%%%%%%%%%%%%%%%%%%%%

We now turn to show that the (bulk) wave functionals with a smooth boundary condition $\phi_- $ defined in (\ref{wavef-g})
are in fact excited CFT states  precisely given by
\be\label{estadoinicial-resultado}
 | \Psi^{\phi_-} \rangle = \, e^{-\int_{\partial_r {\cal M}_-}  \, \mathcal{O}\, \phi_-} \,  |0 \rangle\,,
\ee
provided that $\phi_- =0 $ at $\tau =-\infty$.

For completely arbitrary initial/final states, the prescription (\ref{valoresperadoLorentz}) reads
\be\label{valoresperadoLorentz-psi+-}
 \langle \Psi_f | T[ e^{-i\int_{\partial_r \mathcal{M}_L}  \mathcal{O}\, \phi_L } ]|\Psi_i \rangle =
 \sum_{\phi_{\Sigma^\pm}} \,\,(\Psi_f[\phi_{\Sigma^+}])^*\,\,{\cal Z}[ \phi_L;\phi_{\Sigma^-}, \phi_{\Sigma^+}]\,\, \Psi_i[\phi_{\Sigma^-}] \,.
\ee
Now, let us split the Lorentzian AdS cilinder in figure  \ref{SvR}, whose global time coordinate belongs to the interval $[T_-\, ,\,  T_+]\subset\mathbb{R}$  in two pieces
$\mathcal{M}_L =   \mathcal{M}'_L \cup \tilde{\mathcal{M}} $ joined by the space-like hypersurface $\tilde{\Sigma}$ at $\tilde{T}$ ($T_- \leq \tilde{T} \leq  T_+$). The non trivial boundary condition on its conformal (radial) boundary $\phi_L$ also splits accordingly: $\{\phi'_L,\tilde{\phi} \}$.  Let us also consider $|\Psi^{\phi_-} \rangle \equiv |0 \rangle$, and keep the final state $|\Psi_f \rangle$ arbitrary. The corresponding geometry is shown in figure  \ref{MGP:Fig:ExcStates}, where no geometric dual is associated to the final state
$|\Psi_f \rangle$.

\begin{figure}[t]\centering
\includegraphics[width=.9\linewidth]{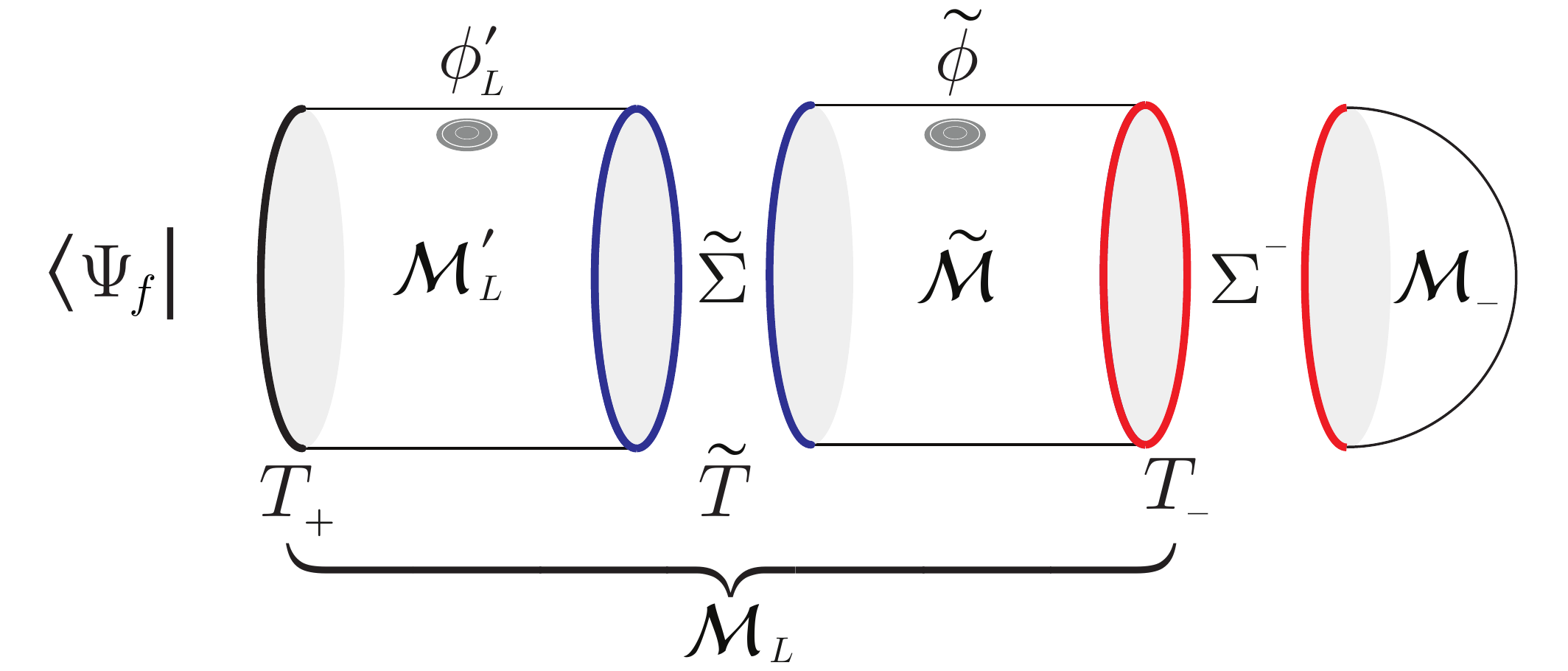}
\caption{The manifold shown can be interpreted as taking figure  \ref{SvR} and splitting it the Lorentzian piece, and also removing
$\mathcal{M}_+$ since in principle, the final state $\Psi_f$ is not necessarily expressed as any (Euclidean) path integral.}\label{MGP:Fig:ExcStates}
\end{figure}

According to the prescription (\ref{SvR-path}), expression (\ref{valoresperadoLorentz-psi+-}) now takes the form
\begin{multline}\label{pre14}
 \langle \Psi_f| T[ e^{- i\int_{\partial_r \mathcal{M}'_L}  \mathcal{O}\, \phi'_L \,\, -i \int_{\partial_r \tilde{\mathcal{M}} }  \mathcal{O}\, \tilde{\phi}}]|0 \rangle = \\ \sum_{\phi_{\tilde\Sigma}, \phi_{\Sigma^\pm}}\,
 (\Psi_f[\phi_{\Sigma^+}])^*
  \left( \int^{\phi_{\Sigma^+}}_{\phi_{\tilde\Sigma}}[\mathcal{D}\Phi]_{\phi'_L} e^{ i S_L[\Phi]}\right)
 \left( \int^{\phi_{\tilde\Sigma}}_{\phi_{\Sigma^-}}[\mathcal{D}\Phi]_{\tilde{\phi}}\,\,e^{ i \tilde{S}[\Phi]} \right)
 \left( \int_{0}^{\phi_{\Sigma^-}} [\mathcal{D}\Phi]_{0}\, e^{- S_-[\Phi]} \right)\,,
\end{multline}
where $\phi_{\tilde\Sigma}$ denotes the value of the field on $\tilde{\Sigma}$.

Now, the key step is to consider a Wick rotation $t\to- i \tau$ only in the piece $\tilde{\mathcal{M}}$. One can then interpret the whole
$\tilde{\mathcal{M}}\cup\mathcal{M}_-$ as describing the \emph{initial state}; in fact, this manifold is completely diffeomorphic to $\mathcal{M}_-$ (a \emph{half} of Euclidean AdS, H). Then,  expression \eqref{pre14} becomes
\be\nn
\langle \Psi_f| T[ e^{- i\int_{\partial_r  \mathcal{M}'_L} \mathcal{O}\, \phi'_L \,\,-\int_{\partial_r \tilde{\mathcal{M}} }  \mathcal{O}\, \tilde{\phi} \,\,}]|0 \rangle
= \sum_{\phi_{\Sigma^+},\phi_{\tilde\Sigma}}\,(\Psi_f[\phi_{\Sigma^+}])^*\,
\left( \int^{\phi_{\Sigma^+}}_{\phi_{\tilde\Sigma}}[\mathcal{D}\Phi]_{\phi'_L} \,\,
   e^{ i S_L[\Phi]}\right)
\left(\int_{0}^{\phi_{\tilde\Sigma}}[\mathcal{D}\Phi]_{\tilde{\phi}}\, e^{-S[\Phi]}\right)\,.
\ee
Finally, taking the limit $|T_+ - \tilde{T} |\to0$, the real time region $\mathcal{M}'_L$ squeezes out, and the surfaces
$\tilde{\Sigma}$ and $\Sigma_+$ coincide, such that $\phi_{\Sigma^+} = \phi_{\tilde\Sigma}$. Thus, the previous equation becomes
\be\label{producto-estadoinicial}
 \langle \Psi_f|  e^{-\int_{\partial_r \tilde{\mathcal{M}} }  \mathcal{O}\, \tilde{\phi} }|0 \rangle =
 \sum_{\phi_{\Sigma^+}}\,\left(\Psi_f[\phi_{\Sigma^+}]\right)^*\left(\int_{0}^{\phi_{\Sigma^+}}[\mathcal{D}\Phi]_{\tilde{\phi}} \,\,
   e^{- S[\Phi]}\,\right)\,.
 \ee
Since this holds \emph{for any} state $|\Psi_{f} \rangle$ at the Hilbert space, and recalling (\ref{identidad}), the initial state can be identified with the ket
\be
 \label{estadoinicial-resultado-CFT}
 |\Psi^{ \phi_-} \rangle \equiv e^{-\int_{\partial_r \tilde{\mathcal{M}}  }  \mathcal{O}\, \tilde\phi }|0 \rangle =
 e^{-\int_{   \partial_r ( \tilde{\mathcal{M}} \cup \mathcal{M}_-) }  \mathcal{O}\, \phi_- }|0 \rangle
= e^{-\int_{\mathcal{S}_-}  \mathcal{O}\, \phi_- }|0 \rangle\,,
\ee
where we have redefined $\mathcal{S}_-\equiv \partial_r(\tilde {\mathcal{M}} \cup \mathcal{M}_{-}) \,- \,\{\tau = - \infty\}$,
and define $\phi_-$ as $0$ over $( -\infty\,,T_-)$, and $\tilde{\phi}$ as $[T_-\, ,\, \tilde{T} )$. Notice that $T_-$ can be chosen arbitrarily close to $-\infty$.
Result \eqref{estadoinicial-resultado-CFT} explicitly shows the relation between the form of the excited state and the boundary conditions
$ \phi_- $ on $\partial_r  \mathcal{M}_-$. In fact, expanding the above exponential in a  Taylor series about the
origin of $\partial_r \mathcal{M}_-$, a combination of primary operators and their descendants arise that act on the vacuum and create CFT excited states.

On the other hand, the projection of the state in the bulk configuration basis is given by the expression within the parenthesis on the rhs of
the equation (\ref{producto-estadoinicial}), which agrees with the proposal (\ref{wavef-g}). This remarkably generalizes the HH method to
asymptotically AdS spacetimes, since it allows to define and evaluate (at least perturbatively on the gravity side) the wave
functional for a family of excited states of the theory characterized by the (radial) boundary data. In fact, we are going to argue below that these excitations shall correspond to
coherent states. In a forthcoming paper, this aspect and its consequences in quantum gravity will be explored more in depth.

%%%%%%%%%%%%%%%%%%%%%%%%%%%%%%%%%%%%%%%%%%%%%%%%%%%%%%%%%%%%%%%%%%%%%%%%%%%%%%%%%%%%%%%%%%%%%%%%%%%%%%%%%%%%%%%%%%%%%%%%%%%%
%%%%%%%%%%%%%%%%%%%%%%%%%%%%%%%%%%%%%%%%%%%%%%%%%%%%%%%%%%%%%%%%%%%%%%%%%%%%%%%%%%%%%%%%%%%%%%%%%%%%%%%%%%%%%%%%%%%%%%%%%%%%
%%%%%%%%%%%%%%%%%%%%%%%%%%%%%%%%%%%%%%%%%%%%%%%%%%%%%%%%%%%%%%%%%%%%%%%%%%%%%%%%%%%%%%%%%%%%%%%%%%%%%%%%%%%%%%%%%%%%%%%%%%%%

\subsection{Quantum coherence from other prescriptions}
\label{coher}

%%%%%%%%%%%%%%%%%%%%%%%%%%%%%%%%%%%%%%%%%%%%%%%%%%%%%%%%%%%%%%%%%%%%%%%%%%%%%%%%%%%%%%%%%%%%%%%%%%%%%%%%%%%%%%%%%%%%%%%%%%%%
%%%%%%%%%%%%%%%%%%%%%%%%%%%%%%%%%%%%%%%%%%%%%%%%%%%%%%%%%%%%%%%%%%%%%%%%%%%%%%%%%%%%%%%%%%%%%%%%%%%%%%%%%%%%%%%%%%%%%%%%%%%%
%%%%%%%%%%%%%%%%%%%%%%%%%%%%%%%%%%%%%%%%%%%%%%%%%%%%%%%%%%%%%%%%%%%%%%%%%%%%%%%%%%%%%%%%%%%%%%%%%%%%%%%%%%%%%%%%%%%%%%%%%%%%

One of our claims in the present work is that, the states (\ref{estadoinicial-resultado-CFT}) prescribed by the SvR framework,
correspond to coherent ones as represented in the gravity/AdS Hilbert space. The following sections will be devoted to compute
time ordered correlation functions between \emph{excited} states following SvR, and check this statement holographically.
Nevertheless if one takes into account other prescriptions for AdS/CFT existing in the literature, and claims consistency with the SvR formalism, this hypothesis can be easily argued.
So in particular, let us briefly recall the recipe \cite{BDHM,BDHM2}, referred to as BDHM.
%Let us briefly explain other prescriptions in the literature \cite{BDHM,BDHM2}, that we will refer as BDHM.
Essentially, if $r\in[0,\infty)$ stands for the global radial coordinate, one canonically quantizes the bulk scalar fields and then identifies the dual
CFT operator with the product $r^{\Delta}\, \Phi$ near the asymptotic boundary, with $\Delta=d/2+\sqrt{d^2/4+m^2}$, $m$ the field $\Phi$ mass and $d$ the CFT dimension.

Consider a classical real scalar field $\Phi$ on $\mathcal{M}_L$ in global coordinates $(t,r,\Omega)$, where $\Omega$ stands for the
angular coordinates in AdS$_{d+1}$ with boundary condition $\phi_L=0$ on $\partial_r\mathcal{M}_L$. It is well known that the general solutions are
  \begin{equation}\nonumber
\Phi(t,r,\Omega) = \sum_k  \Big( a_k^*  f^*_k(t,r,\Omega) + a_k f_k(t,r,\Omega) \Big)\,,
\end{equation}
where $f_k(t,r,\Omega)$ stands for the positive frequency normalizable modes, fixed such that satisfy the orthonormality relations
\begin{equation}\nonumber
\left( f_k \,, f_{k'} \right) = \delta_{k k'}\,,
\end{equation}
where $k$ schematically denotes all the numbers that label a particular positive-frequency solution and $(\quad,\quad)$ is the Klein-Gordon product on AdS$_{d+1}$. Thus, the coefficients $a_k^*$, and $a_k$ can be promoted to operators by
canonical quantization as \cite{avis}
\begin{equation}\nonumber
\hat{\Phi }(t,r,\Omega) = \sum_k  \Big( \hat{a}_k^\dagger f^*_k(t,r,\Omega) + \hat{a}_k f_k(t,r,\Omega)\Big)\,.
\end{equation}
Therefore, the BDHM prescription identifies the dual CFT operators as \cite{BDHM},\cite{kaplan}
\be
\label{kaplan}
\hat{ \mathcal{O}}(t,\Omega) \equiv\lim _{r\to\infty} \,r^{\Delta}\,\hat{\Phi}(t,r,\Omega)  =
\sum_k  \hat{a}_k^\dagger F^*_k(t,\Omega) + \hat{a}_k F_k(t,\Omega)\,,
\ee
which defines a basis of functions on the conformal boundary
\begin{equation}
\label{eigenf-boundarygen}
F_k(t,\Omega) \equiv \lim _{r\to\infty} r^{\Delta}f_k(t,r,\Omega) = N_k\, e^{-i \omega_k t}Y_k(\Omega)\,\, .
\end{equation}
Here $Y_k(\Omega)$ stand for the spherical harmonics on $S^{d-1}$, and $N_k$ are numeric factors\footnote{For instance, for AdS$_{2+1}$ in global coordinates one finds
\be\label{eigenf-boundary}
F_{nl}(t,\varphi)   = \sqrt{\frac{\Gamma[\Delta+n+|l|]\Gamma[\Delta+n]}{n! (\Gamma[\Delta])^2\Gamma[n+|l|+1]}}
\frac{1}{\sqrt{2\pi}} e^{-i \omega_{nl} t+il\varphi}\quad\Longrightarrow\quad N_{nl}=\sqrt{\frac{\Gamma[\Delta+n+|l|]\Gamma[\Delta+n]}{n! (\Gamma[\Delta])^2\Gamma[n+|l|+1]}}.
\ee    }.

Finally, if we demand consistency of \eqref{estadoinicial-resultado-CFT} with \eqref{kaplan}, we
conclude that the SvR state \eqref{estadoinicial-resultado-CFT} is \emph{coherent} and can be written as
 \begin{equation}\label{coherente-st}
 |\Psi^{{\phi}_- } \rangle  \propto e^{\sum_k \, \lambda^-_k \, \hat{a}_k^\dagger} |0 \rangle\,,
\end{equation}
where
\be
\label{lambdak}
\lambda^\pm_k = -\int_{\partial_r {\cal M}_-}  \, d\tau d\Omega \, F^*_k(-i\tau,\Omega) \,\phi_\pm(\tau,\Omega)\,,
\ee
which  requires the analytical extension of the functions (\ref{eigenf-boundarygen}) to
\be\nn
F^*_k(-i\tau,\Omega) = N_k \, e^{ \omega_k \tau} \,\,Y_k(\Omega)\,.
\ee%\label{eigenf-boundaryE}
This is what will be checked in the following sections through explicit holographic SvR computations of  one- and two-points
correlation functions. In fact, since coherent states \eqref{coherente-st} are eigenstates of the annihilation operators
\be\nn \hat{a}_k |\Psi^{\phi_\pm} \rangle = \lambda^\pm_k |\Psi^{\phi_\pm} \rangle \,, \ee %\label{autovalores-k}
we are able to compute the eigenvalues $\lambda^\pm_k$ and then construct the state (\ref{coherente-st}) explicitly.
Precisely, by virtue of (\ref{kaplan}) and (\ref{coherente-st}), we can notice the following useful relation
\be\label{kaplan2}
\frac{\langle  \Psi^{\phi_+}| \mathcal{O}(t,\Omega) |\Psi^{\phi_-} \rangle}{\langle \Psi^{\phi_+} | \Psi^{\phi_-} \rangle} = \sum_k  (\lambda^+_k )^{*}\,\, F^*_k(t,\Omega) + \lambda^-_k\,\, F_k(t,\Omega)\,\,,
\ee
which will allow us to identify the eigenvalues computed holographically and to test this construction.

Let us end this section by stressing that expression (\ref{kaplan}) results from the fact that we are considering a free theory on the bulk, which according to the AdS/CFT dictionary, implicitly assumes the large N limit (see ref. \cite{BDHM} for a more detailed discussion).

%%%%%%%%%%%%%%%%%%%%%%%%%%%%%%%%%%%%%%%%%%%%%%%%%%%%%%%%%%%%%%%%%%%%%%%%%%%%%%%%%%%%%%%%%%%%%%%%%%%%%%%%%%%%%%%%%%%%%%%%%%%%
%%%%%%%%%%%%%%%%%%%%%%%%%%%%%%%%%%%%%%%%%%%%%%%%%%%%%%%%%%%%%%%%%%%%%%%%%%%%%%%%%%%%%%%%%%%%%%%%%%%%%%%%%%%%%%%%%%%%%%%%%%%%%
%%%%%%%%%%%%%%%%%%%%%%%%%%%%%%%%%%%%%%%%%%%%%%%%%%%%%%%%%%%%%%%%%%%%%%%%%%%%%%%%%%%%%%%%%%%%%%%%%%%%%%%%%%%%%%%%%%%%%%%%%%%%%

\subsection{The inner product between asymptotic states in the SvR prescription, and first holographic check of coherence}
\label{innerprod}
%%%%%%%%%%%%%%%%%%%%%%%%%%%%%%%%%%%%%%%%%%%%%%%%%%%%%%%%%%%%%%%%%%%%%%%%%%%%%%%%%%%%%%%%%%%%%%%%%%%%%%%%%%%%%%%%%%%%%%%%%%%
%%%%%%%%%%%%%%%%%%%%%%%%%%%%%%%%%%%%%%%%%%%%%%%%%%%%%%%%%%%%%%%%%%%%%%%%%%%%%%%%%%%%%%%%%%%%%%%%%%%%%%%%%%%%%%%%%%%%%%%%%%%%
%%%%%%%%%%%%%%%%%%%%%%%%%%%%%%%%%%%%%%%%%%%%%%%%%%%%%%%%%%%%%%%%%%%%%%%%%%%%%%%%%%%%%%%%%%%%%%%%%%%%%%%%%%%%%%%%%%%%%%%%%%%%

Noticeably, the SvR prescription allows to compute the inner product between arbitrary initial/final states. In fact, if one
squeezes the Lorentzian manifold ${\cal M}_L$ by taking the limit $|T_+ - T_- |\to0$ in  (\ref{valoresperadoLorentz-psi+-}),
one obtain a remarkable holographic formula for the inner product of (in/out) CFT-states
\be
\label{inprod}
 \langle \Psi^{\phi_+} | \Psi^{\phi_-} \rangle = \sum_{\phi_\Sigma} \,\,
 \left(\Psi^{\,\phi_+}[\phi_\Sigma]\,\right)^*\,\,\, \Psi^{\phi_-}[ \phi_\Sigma] \,= {\cal Z}_E[\phi_-, \phi_+]\,,
\ee
where $\phi_\Sigma$ is the value of $\Phi$ on a hypersurface surface $\Sigma$ embedded into Euclidean AdS,
that intersects the boundary at the \emph{equator} that divides the $S^{d}$ sphere into two hemispheres $\partial_r  \mathcal{M}_\pm$, and
$\phi_\pm$ denotes the respective boundary values, as shown in figure 
\ref{Fig:Bola}. In the saddle point approximation \eqref{inprod} reads
\begin{figure}\centering
\includegraphics[width=.5\linewidth]{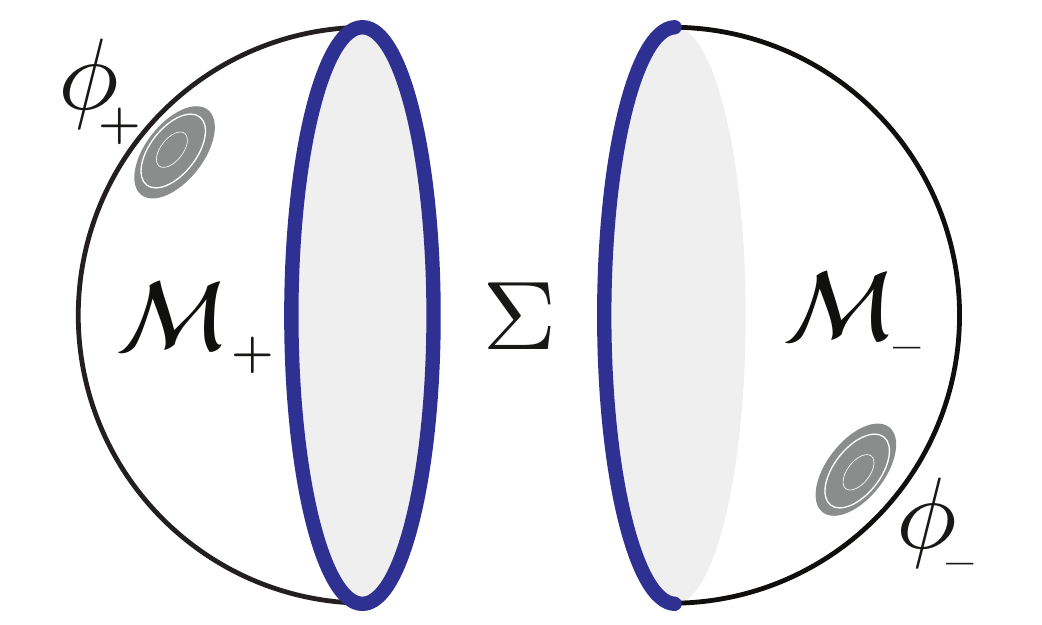}
\caption{The configuration considered in figure  \ref{SvR}b becomes an Euclidean AdS space when the Lorentzian cilinder ${\cal M}$ length is set to zero.}\label{Fig:Bola}
\end{figure}
\be\nn
 \langle \Psi^{\phi_+} | \Psi^{\phi_-} \rangle = e^{- S^0_E[\phi_+,\phi_-]} = e^{-\int_{\partial_r  \mathcal{M}_-} \sqrt{|\gamma_r|}\phi_-\, \, n^\mu\nabla_\mu \Phi(\phi_\pm)  \,-\,\int_{\partial_r  \mathcal{M}_+} \sqrt{|\gamma_r|} \phi_+\, \, n^\mu\nabla_\mu \Phi(\phi_\pm)}.
 \ee%\label{inner-product}
Then, using the well known result \cite{GKP1,GKP2}, we find and holographic expression for the inner product
\begin{align}
 \langle \Psi^{\phi_+} | \Psi^{\phi_-} \rangle &= e^{- S^0_{E}[\phi_+,\phi_-]} = e^{-\int_{S^{d}}dx\, \int_{S^{d}} dy \,\phi(x) \, G(x,y) \,\phi (y) } \, \nn \\
 &= e^{- \left\{ \int_{\mathcal{S}_-} \, \int_{\mathcal{S}_-}\, \phi_-(x) \, G(x,y) \,\phi_- (y)
  +  \int_{\mathcal{S}_+}\int_{\mathcal{S}_+}\,dx dy\, \phi_+(x) \, G(x,y) \,\phi_+ (y)
   +\, 2 \int_{\mathcal{S}_-}\int_{\mathcal{S}_+}\,dx dy\, \phi_-(x) \, G(x,y) \,\phi_+ (y)\right\}}\,,
   \label{botas}
 \end{align}
where $G$ is the boundary-to-boundary Green function defined in \cite{GKP2}, $\phi$ is defined on the (open) intervals
$ {\cal S}_- \equiv (-\infty , 0) \times S^{d-1} $ and $ {\cal S}_+ \equiv (0 , \infty ) \times S^{d-1}$ by the smooth functions
$\{ \phi_-(x) \,, \,\phi_+ (x)\}$ respectively, and the points $x, y$ are parameterized by coordinates $(\tau, \Omega) \,\in\, I\!\!R \times S^{d-1}$.
Then, one can use this formula in particular to compute the norm of states (\ref{estadoinicial-resultado}), which in principle are not normalized.
However we shall first define the corresponding (conjugate) dual $\langle \Psi^{\phi_-} |$.
Since the two hemispheres $\mathcal{S}_\pm$ are diffeomorphic, one can extend the definition of functions (and operators) into each other through the dualization map\footnote{Therefore, expressions as \eqref{inprod} should be interpreted as 
\begin{equation*}
 \left( \Psi^{\,\phi_+} ( \phi ) \right)^* 
 \equiv  \Psi^{\,\phi^\star_+} ( \phi ).
 \end{equation*}}
\be
\label{conjugate}
\phi^\star_\pm (\tau, \Omega) \equiv \phi_\pm (-\tau , \Omega ) .
\ee
 This is the standard conjugation prescription in Euclidean CFT theories \cite{jackiw}, where, in the setup known as radial quantization, the
 ``radius'' $r= e^{-\tau}$ maps to $r^{-1}= e^{\tau}$ (and $\tau \mapsto -\tau$).

The norm of an arbitrary state, say $| \Psi^{\phi_-} \rangle$, is therefore given by
\be
\langle \Psi^{\phi_-} | \Psi^{\phi_-} \rangle = \|\Psi^{\phi_-}\|^2 =  e^{- S^0_E[\phi^\star_-,\phi_-]} \,,
\ee%\label{norm}
where, using (\ref{conjugate})
\begin{align}
S^0_{E}[\phi^\star_-,\phi_-] =& \int^0_{-\infty}d\tau \, \int^0_{-\infty} d\tau' \,\, \phi_-(\tau) \, G(\tau , \tau') \,\phi_- (\tau')
+ \int_0^{\infty}d\tau \, \int_0^{\infty} d\tau' \,\, \phi_-(-\tau) \, G(\tau, \tau') \,\phi_- (-\tau') \nn\\
&+\,\, 2 \int^0_{-\infty}d\tau \, \int_0^{\infty} d\tau' \,\, \phi_-(\tau) \, G(\tau, \tau') \,\phi_- (-\tau') \nn \\
=& 2 \int^0_{-\infty}d\tau \, \int^0_{-\infty} d\tau' \,\, \phi_-(\tau) \, G(\tau , \tau') \,\phi_- (\tau') +
\,\, 2 \int^0_{-\infty}d\tau \, \int_0^{\infty} d\tau' \,\, \phi_-(\tau) \, G(\tau, \tau') \,\phi_- (-\tau')\nn\,.
\end{align}%\label{GKPW2-2}
The integrals on the angular variables $\Omega, \Omega' $ are implicit to simplify the notation. To find this result we have also used the properties of the boundary-to-boundary Green function $G(-\tau, -\tau') = G(\tau, \tau')$ and $G(\tau, \tau') = G(\tau', \tau)$.

So therefore, the inner product between two \emph{normalized} excited states defined within the SvR formulae reads
\be
\langle \Psi_{_{\cal N}}^{\phi_+} | \Psi_{_{\cal N}}^{\phi_-} \rangle =  \frac{1}{\|\Psi^{\phi_+}\|\|\Psi^{\phi_-}\| }e^{- S^0_{E}[\phi_+,\phi_-]} =
e^{- \left( S^0_{E}[\phi_+,\phi_-]- \frac{1}{2 } S^0_{E}[\phi_+,\phi^\star_+]- \frac{1}{2}S^0_{E}[\phi^\star_-,\phi_-]\right)}\,,
\ee %\label{producto-ests-normalizados}
where the exponent
\begin{multline}\nn
S^0_{E}[\phi_+,\phi_-]- \frac{1}{2 } S^0_{E}[\phi_+,\phi^\star_+]- \frac{1}{2}S^0_{E}[\phi^\star_-,\phi_-] =\\
2\int^0_{-\infty}d\tau \, \int_0^{\infty} d\tau' \Bigg( \phi_-(\tau) \, G(\tau , \tau') \,\phi_+ (\tau')
- \frac 12 \phi_+(\tau) \, G(\tau, \tau') \,\phi_+ (-\tau') - \frac 12 \phi_-(\tau) \, G(\tau, \tau') \,\phi_- (-\tau') \Bigg)\,,
\end{multline}
can be regrouped as
\begin{multline}\label{prod-reagrup}
-\int^0_{-\infty}d\tau \, \int_0^{\infty} d\tau' \,\, \left(\phi_-(\tau) - \phi_+(-\tau) \right)\, G(\tau, \tau') \, \left(\phi_-(-\tau') - \phi_+(\tau')\right)=\\
\int^0_{-\infty}d\tau \, \int^0_{\infty} d\tau' \,\, \left(\phi_-(\tau) - \phi^\star_+(\tau) \right)\, G(\tau, \tau') \, \left(\phi_-(\tau') - \phi_+^\star(\tau')\right)^\star\,.
\end{multline}
This expression can be written as $\left( ( \phi_- \, - \, \phi^\star_+)\, , \,( \phi_- \, - \, \phi^\star_+\,) \right)=| \phi_- \, - \, \phi^\star_+ |^2 $ 
if we define the inner product
\be\label{product-final-reales}
(\phi_1 \, ,\, \phi_2) \equiv \int^0_{-\infty}d\tau \, \int^0_{\infty} d\tau' \,\, \phi_1(\tau)\, G(\tau, \tau') \, \phi_2^\star(\tau')\,,
\ee
on the space of functions $\mathcal{C}\equiv\{\phi_\pm  \,,\phi_1 \,,\phi_2 \,,\,\dots\}$ defined on one of the hemispheres $-\infty<\tau < 0$ that parameterize the CFT states. Note that this product is given by a non-degenerate metric on this space of functions, whose matrix elements are given by the (boundary-to-boundary) two points function.
Let us observe that in the second line of (\ref{prod-reagrup}) we have written the integral in a symmetric way,
 on the interval  $(\infty , 0) \times S^{d-1} $ in place of $\mathcal{S}_+$, which properly expresses the integration as ordered from the (north) pole to the boundary of $\mathcal{S}_+$ (equator).

This result constitutes the first holographic check of our claim on the coherence of the in/out states in the SvR proposal, since the product of two (normalized) states can indeed be written in terms of the norm, induced by a scalar product on the space of states as
\be\label{final-product}
 \langle \Psi_{_{\cal N}}^{\,\phi_+} | \Psi_{_{\cal N}}^{\,\phi_-} \rangle = e^{-| \phi_-\, -\, \phi^\star_+|^2}\,.
\ee
Reinserting the $1/G_N\sim N^2$ factor in front of the sugra scalar action \eqref{botas}, one finds that the Gaussian width in 
\eqref{final-product} is controlled by $N$. It is immediate to see that the rhs of 
\eqref{final-product} becomes localized in $\mathcal{C}$ as $ N\to \infty$, as expected for (semi-) classical states.

%%%%%%%%%%%%%%%%%%%%%%%%%%%%%%%%%%%%%%%%%%%%%%%%%%%%%%%%%%%%%%%%%%%%%%%%%%%%%%%%%%%%%%%%%%%%%%%%%%%%%%%%%%%%%%%%%%%%%%%%%%%%
%%%%%%%%%%%%%%%%%%%%%%%%%%%%%%%%%%%%%%%%%%%%%%%%%%%%%%%%%%%%%%%%%%%%%%%%%%%%%%%%%%%%%%%%%%%%%%%%%%%%%%%%%%%%%%%%%%%%%%%%%%%%%
%%%%%%%%%%%%%%%%%%%%%%%%%%%%%%%%%%%%%%%%%%%%%%%%%%%%%%%%%%%%%%%%%%%%%%%%%%%%%%%%%%%%%%%%%%%%%%%%%%%%%%%%%%%%%%%%%%%%%%%%%%%%%

Although in this paper we work with real scalar fields, let us finally comment that for complex ones $\phi \equiv \phi^R + i \phi^I $,
the action decouples into two independent terms and the derivation above follows straightforwardly for each component field, such that
regroups as a sum of two terms as (\ref{prod-reagrup}) for each component,
\be
  \left( ( \phi^R_- \, - \, (\phi_+^R)^\star)\, , \,( \phi^R_- \, - \, (\phi_+^R)^\star\,) \right) +
  \left( ( \phi^I_- \, - \, (\phi_+^I)^\star)\, , \,( \phi^I_- \, - \, (\phi_+^I)^\star\,) \right)\,,
\ee
with the inner product $(\;,\;)$ rule as given in (\ref{product-final-reales}) and the rule (\ref{conjugate}) assumed separately for both $\phi^{R/I}$.
 Thus, the final result can be written as
  \be\label{final-product-complex}
 \langle \Psi_{_{\cal N}}^{\,\phi_+} | \Psi_{_{\cal N}}^{\,\phi_-} \rangle = e^{-\left(| \phi^R_-\, -\, \phi^{R\star}_+ |^2 \,+\,| \phi^I_-\, -\, \phi^{I\star}_+|^2\right) }\,.
\ee
Therefore, for the space of complex fields defined on  $\mathcal{S}_-$ the product (\ref{product-final-reales}) can naturally be generalized to
\be
(\phi_1 \, ,\, \phi_2) \equiv   \,\int^0_{-\infty}d\tau \, \int^0_{\infty} d\tau' \,\, \phi_1(\tau)\, G(\tau, \tau') \, (\phi_2^{\star}(\tau'))^*\,,
\ee
where $( )^*$ stands for the standard complex conjugation operation.

%%%%%%%%%%%%%%%%%%%%%%%%%%%%%%%%%%%%%%%%%%%%%%%%%%%%%%%%%%%%%%%%%%%%%%%%%%%%%%%%%%%%%%%%%%%%%%%%%%%%%%%%%%%%%%%%%%%%%%%%%%%%
%%%%%%%%%%%%%%%%%%%%%%%%%%%%%%%%%%%%%%%%%%%%%%%%%%%%%%%%%%%%%%%%%%%%%%%%%%%%%%%%%%%%%%%%%%%%%%%%%%%%%%%%%%%%%%%%%%%%%%%%%%%%%
%%%%%%%%%%%%%%%%%%%%%%%%%%%%%%%%%%%%%%%%%%%%%%%%%%%%%%%%%%%%%%%%%%%%%%%%%%%%%%%%%%%%%%%%%%%%%%%%%%%%%%%%%%%%%%%%%%%%%%%%%%%%%

\section{Expectation values of local operators}
\label{ExpVal}

%%%%%%%%%%%%%%%%%%%%%%%%%%%%%%%%%%%%%%%%%%%%%%%%%%%%%%%%%%%%%%%%%%%%%%%%%%%%%%%%%%%%%%%%%%%%%%%%%%%%%%%%%%%%%%%%%%%%%%%%%%%%
%%%%%%%%%%%%%%%%%%%%%%%%%%%%%%%%%%%%%%%%%%%%%%%%%%%%%%%%%%%%%%%%%%%%%%%%%%%%%%%%%%%%%%%%%%%%%%%%%%%%%%%%%%%%%%%%%%%%%%%%%%%%%
%%%%%%%%%%%%%%%%%%%%%%%%%%%%%%%%%%%%%%%%%%%%%%%%%%%%%%%%%%%%%%%%%%%%%%%%%%%%%%%%%%%%%%%%%%%%%%%%%%%%%%%%%%%%%%%%%%%%%%%%%%%%%

In this section we generalize the SvR computation \cite{SvRC} to the case of excited states obtained by turning on
boundary conditions in the Euclidean sections as discussed before (see figure \ref{SvR}). Our aim is to characterize the   excited state.

We will work with the simplest model, a real massive scalar field $\Phi$  in AdS$_{d+1}$, and for the ease of computations we will
consider the semiclassical approximation and $d=2$. This setup allows to compute correlation functions of dual local
scalar operator  $\mathcal{O}$  with conformal dimension $\Delta=d/2 + \sqrt{d^2/4+m^2}$, with $m$ being the mass, in terms of classical bulk solutions.

At the formal level the scalar field action
\begin{equation}
\label{ficmplx}
S[\Phi] = -\frac{1}{2} \int_{C} d{\mathsf t}  \int dr d\varphi \sqrt{|g|}\left(\partial_{\mu} \Phi  \partial^{\mu} \Phi + m^2 \Phi\right)\,,	
\end{equation}
is defined over the contour $C$ on the complex $\sf t$-plane
shown in figure \ref{SvR}a \cite{SvRC} (see \cite{SS} for related work).
Without loss of generality we can take $T_{\pm}\equiv \pm T$. Calling $\Phi_-(\tau)  \equiv \Phi(-T-i\tau)$, with $\tau\in(-\infty,0] $  the field on the left  vertical piece,
$\Phi_L(t) \equiv \Phi(t),~ t\in[-T,T]$ the field on the horizontal piece and $\Phi_+(\tau)  \equiv \Phi(T-i\tau),~\tau\in[0,\infty) $ the field on the right vertical piece,
equation \eqref{ficmplx} becomes
%\footnote{The contour $C$ can be parametrized as $$ {\sf t}(\lambda)=\left\{\begin{array}{cc} -T -i(\lambda+T) & -\infty<\lambda<-T  \\ \lambda & -T<\lambda<T\\ T -i(\lambda-T ) & T <\lambda<\infty \end{array}\right., $$ the gluing \eqref{glue} taking place at points ${\sf t}= \pm T$ in the complex $\sf t$-plane.}
\be
\nn
S =i S_- + S_L + i S_+\,,
\ee
where
\begin{align}
S_\pm =&+\frac 12 \int_{\mathcal{M}_\pm} dx^{d+1} \sqrt{|g|} \left(\partial_{\mu} \Phi_\pm \partial^{\mu} \Phi_\pm  + m^2  \Phi_\pm ^2\right)\nn \\=&- \frac 12 \int_{\mathcal{M}_\pm} dx^{d+1} \sqrt{|g|}\,\, \Phi_\pm  \left( \Box - m^2\right) \Phi_\pm \nn \\&
+ \frac{1}{2} \int_{\partial_r\mathcal{M}_\pm} dx^{d} \sqrt{|\gamma_r|}\,\, \Phi_\pm  n_r^{\mu} \partial_{\mu}\Phi_\pm + \frac 12 \int_{\Sigma^\pm} dx^{d} \sqrt{|\gamma_{_\Sigma}|}\,\, \Phi_\pm  n_{_\Sigma}^{\mu} \partial_{\mu}\Phi_\pm \label{SE}\,,
\end{align}
and
\begin{align}
S_L =& -\frac12\int_{\mathcal{M}_L} dx^{d+1} \sqrt{|g|} \left(\partial_{\mu} \Phi_L \partial^{\mu} \Phi_L  + m^2  \Phi_L ^2\right) \nn \\=&+ \frac 12 \int_{\mathcal{M}_L} dx^{d+1} \sqrt{|g|}\,\, \Phi_L  \left( \Box - m^2\right) \Phi_L   -\frac{1}{2} \int_{\partial_r\mathcal{M}_L} dx^{d} \sqrt{|\gamma_r|}\,\, \Phi_L  n_r^{\mu} \partial_{\mu}\Phi_L \nn \\&
- \frac 12 \int_{\Sigma^+} dx^{d} \sqrt{|\gamma_{_\Sigma}|}\,\, \Phi_L n_\Sigma^{\mu} \partial_{\mu}\Phi_L - \frac 12 \int_{\Sigma^-} dx^{d} \sqrt{|\gamma_{_\Sigma}|}\,\, \Phi_L n_\Sigma^{\mu} \partial_{\mu}\Phi_L \label{SL}
\,.
\end{align}
In these expressions an integration by parts led to boundary terms involving outer normal vectors to the bulk $n^\mu$.
These terms are of two distinct types: conformal boundaries denoted either by $r$ or $\partial_r$ (boundaries at infinity) and $\Sigma$-type spacelike
boundaries as shown in figure \ref{SvR}b\footnote{A possible boundary contribution arising from $\Sigma$ hypersurfaces at  $\tau\to\pm\infty$ in
\eqref{SE} is absent since, as discussed in section \ref{construct}, the boundary conditions turn off at these points.}. Finally, $\gamma_r$ and $\gamma_{_\Sigma}$ are the induced
metric determinants over the corresponding surfaces, and $\Box\phi\equiv(\sqrt{|g|})^{-1}\partial_{\mu}(\sqrt{|g|}g^{\mu\nu}\partial_{\nu}\phi)$
is the standard scalar Laplacian.

The task is to find a smooth classical solution to \eqref{ficmplx} with $\phi_\pm,\phi_L\neq 0$.
In order to find such configuration, as explained in Sec. \ref{construct},  we will solve the appropriate field equation on each section and impose continuity of $\Phi $ and
its conjugated momentum at ${\sf t}=\pm T$. Explicitly, we shall demand
\begin{align}
\Phi_-(r,0,\varphi)&=\Phi_L(r,-T,\varphi)\,,  & \Phi_L(r,T,\varphi)&=\Phi_+(r,0,\varphi)\,, \nn \\
\partial_{\tau} \Phi_-(r,\tau,\varphi)\Big|_{\tau=0}  &= -i \partial_{t} \Phi_L(r,t,\varphi)\Big|_{t=-T}\,,
& -i \partial_{t} \Phi_L(r,t,\varphi)\Big|_{t=T} &= \partial_{\tau} \Phi_+(r,\tau,\varphi)\Big|_{\tau=0}\,. \label{Exp:InOut:CC}
\end{align}
As it is standard \cite{GKP1}, the problem must be regularized by imposing the boundary conditions $\phi_\pm,\phi_L$ at a finite radial cut-off $R$,
at the end the limit $R\to\infty$ is taken.

%%%%%%%%%%%%%%%%%%%%%%%%%%%%%%%%%%%%%%%%%%%%%%%%%%%%%%%%%%%%%%%%%%%%%%%%%%%%%%%%%%%%%%%%%%%%%%%%%%%%%%%%%%%%%%%%%%%%%%%%%%%%
%%%%%%%%%%%%%%%%%%%%%%%%%%%%%%%%%%%%%%%%%%%%%%%%%%%%%%%%%%%%%%%%%%%%%%%%%%%%%%%%%%%%%%%%%%%%%%%%%%%%%%%%%%%%%%%%%%%%%%%%%%%%%
%%%%%%%%%%%%%%%%%%%%%%%%%%%%%%%%%%%%%%%%%%%%%%%%%%%%%%%%%%%%%%%%%%%%%%%%%%%%%%%%%%%%%%%%%%%%%%%%%%%%%%%%%%%%%%%%%%%%%%%%%%%%%

\subsection{Solution  over $\mathcal{M}_L$}

%%%%%%%%%%%%%%%%%%%%%%%%%%%%%%%%%%%%%%%%%%%%%%%%%%%%%%%%%%%%%%%%%%%%%%%%%%%%%%%%%%%%%%%%%%%%%%%%%%%%%%%%%%%%%%%%%%%%%%%%%%%%
%%%%%%%%%%%%%%%%%%%%%%%%%%%%%%%%%%%%%%%%%%%%%%%%%%%%%%%%%%%%%%%%%%%%%%%%%%%%%%%%%%%%%%%%%%%%%%%%%%%%%%%%%%%%%%%%%%%%%%%%%%%%%
%%%%%%%%%%%%%%%%%%%%%%%%%%%%%%%%%%%%%%%%%%%%%%%%%%%%%%%%%%%%%%%%%%%%%%%%%%%%%%%%%%%%%%%%%%%%%%%%%%%%%%%%%%%%%%%%%%%%%%%%%%%%%

For the Lorentzian region we write the AdS metric in global coordinates as (setting the AdS radius ${\cal R}=$1)
\begin{equation*}
ds^{2} = -(1+r^{2})dt^{2} + \frac{dr^{2}}{1+r^{2}}+r^{2}d\varphi^{2}\,.
\end{equation*}
Plugging the ansatz $\Phi_{L}(r,t,\varphi) \propto e^{-i \omega t+i l \varphi}\, f(\omega,l,r)$ with $l\in\mathbb{Z}$, into the
Klein Gordon (KG) equation following from \eqref{SL} gives
\begin{equation}\nn
\left(\frac{1}{r}\partial_{r}\Big(r(1+r^{2})\partial_{r}\Big) + \frac{\omega^{2}}{1+r^{2}} - \frac{l^{2}}{r^{2}} - m^{2} \right) f(\omega,l,r) = 0\,.
\end{equation}
%\label{Exp:InOut:EcDifLorentz}
The regular solution to this equation is
\begin{equation}\nn
%\label{KGsolL}
f(\omega,l,r) = (1+r^{2})^{\sqrt{\omega^{2}}/2}\,\, r^{|l|}\,\,
_2F_1\left( \frac{\sqrt{\omega^{2}}+|l|+\Delta}{2}, \frac{\sqrt{\omega^{2}}+|l|-\Delta+2}{2} ; 1+|l| ; -r^{2}\right),
\end{equation}
where $_2F_1(a,b;c;x)$ is Gauss hypergeometric function. A general solution to the KG equation can be obtained as
\be
\Phi_L(r,t,\varphi)=\sum_{l\in\mathbb{Z}} \int d\omega\, c_{\omega l}\, e^{-i \omega t + i l \varphi}  f(\omega,l,r)\,.
\label{Lc}
\ee
The regularized boundary condition to be imposed is
\begin{equation}
\label{Exp:InOut:aExp}
\Phi_{L}(R,t,\varphi)=R^{\Delta-2} \phi_L(t,\varphi).
\end{equation}
From \eqref{Lc} and \eqref{Exp:InOut:aExp} one obtains
\begin{equation}
\nn
c_{\omega l}= \frac{R^{\Delta-2}}{4\pi^2 } \int dt' d\varphi' e^{i \omega t' - i l \varphi'}\frac{ \phi_L(t',\varphi')}{f(\omega,l,R)}\,.
\end{equation}
Inserting this expression in \eqref{Lc} one finds
\begin{equation}
\label{svrfi}
\Phi_{L}(r,t,\varphi)=\frac{R^{\Delta-2}}{4\pi^2 } \sum_{l\in\mathbb{Z}} \int d\omega  dt' d\varphi' e^{-i \omega (t-t') + i l (\varphi-\varphi')}  \phi_L(t',\varphi') \frac{f(\omega,l,r)}{f(\omega,l,R)}\,.
\end{equation}
Notice that this expression is ill defined since $f(\omega,l,R)=0$ for $\omega=\pm\omega_{nl}^{_R}$ such that
\be
\label{modeR}
\omega_{nl}^{_R}\equiv \omega_{nl} + \epsilon(R)\,,
\ee
where $\omega_{nl}=2n+|l|+\Delta$,  ($n=0,1,2,\ldots$), and $\epsilon(R)\sim {  o}(1/R^{2\Delta-2})$. Modes \eqref{modeR} become the standard Dirichlet AdS normalizable modes in the $R\to\infty$ limit. Following \cite{SvRC} we give meaning to \eqref{svrfi} by choosing the Feynman
path on the complex $\omega$-plane as shown in figure \ref{Exp:Fig:PolosA}. As a consequence the general solution to the KG
equation satisfying \eqref{Exp:InOut:aExp} is given by
\begin{align}
\Phi_L(r,t,\varphi)= &\frac{R^{\Delta-2}}{4\pi^2 } \sum_{l\in\mathbb{Z}} \int_{\mathcal{F}} d\omega dt' d\varphi' e^{-i \omega (t-t') + i l (\varphi-\varphi')}  \phi_L(t',\varphi') \frac{f(\omega,l,r)}{f(\omega,l,R)} \nn\\
&+ \sum_{\substack{n\in\mathbb{N}\\ l\in\mathbb{Z} }}\Big(
 L_{nl}^{+}\, e^{- i \omega_{nl}^{_R} t}+ L_{nl}^{-}\, e^{+ i \omega_{nl}^{_R} t}\Big)e^{  i l \varphi}  g_{nl}(r)\,,
 \label{Exp:InOut:Sol}
\end{align}
where the $L_{nl}^{\pm}$ coefficients will be determined once we impose boundary conditions \eqref{Exp:InOut:CC} at
$\Sigma^\pm$-hypersurfaces (see \eqref{Coef}), and
\begin{equation}
\label{gnl}g_{nl}(r)\equiv f(\pm\omega_{nl}^{_R},l,r) \,.
\end{equation}

\begin{figure}[t]\centering
\begin{subfigure}{0.48\textwidth}\centering
\includegraphics[width=.9\linewidth]{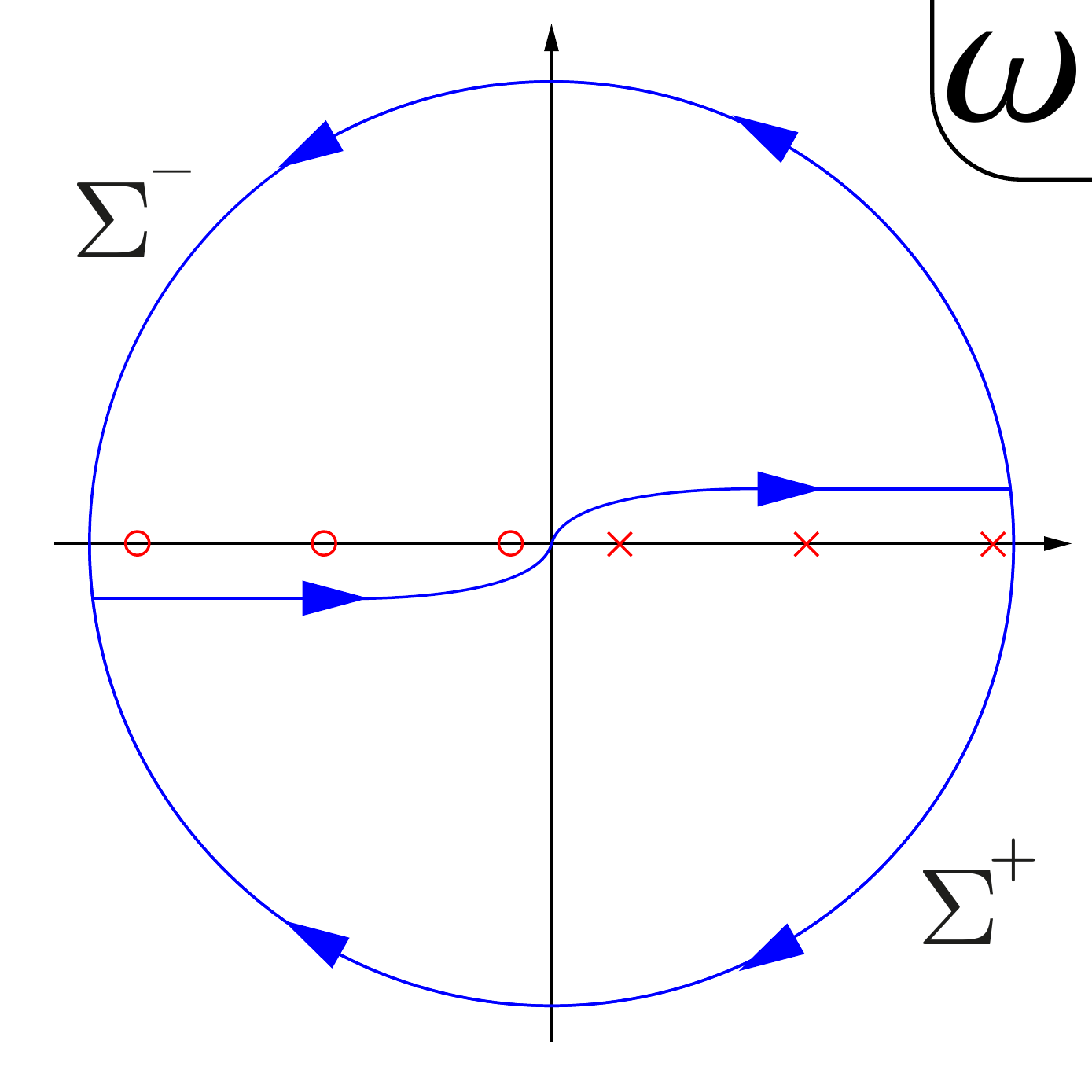}
\caption{}\label{Exp:Fig:PolosA}
\end{subfigure}
\begin{subfigure}{0.48\textwidth}\centering
\includegraphics[width=.9\linewidth]{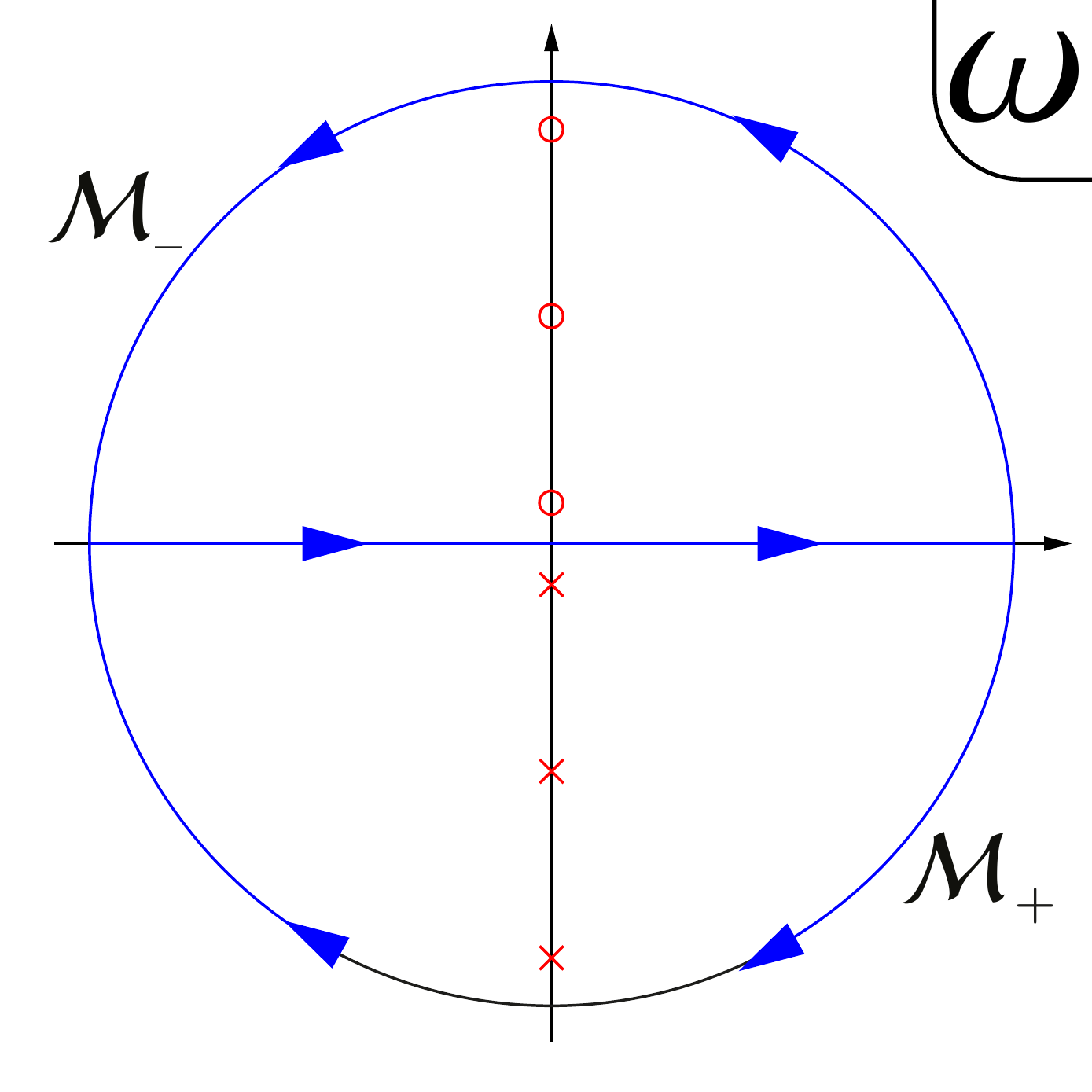}
\caption{}\label{Exp:Fig:PolosB}
\end{subfigure}
\caption{(a) Feynman path $\cal F$ in the complex $\omega$-plane chosen to define (\ref{Exp:InOut:Sol}). $\Sigma^\pm$ indicate the
appropriate contour one should choose near each hypersurface to compute the integral by residues. (\subref{Exp:Fig:PolosB})
$\omega$-plane location of zeroes of $f(-i\omega,l,R)$ appearing  in \eqref{Exp:InOut:ENNSol}. ${\cal M}_\pm$ represent the
appropriate integration path for residues computation .}\label{Exp:Fig:Polos}
\end{figure}

%%%%%%%%%%%%%%%%%%%%%%%%%%%%%%%%%%%%%%%%%%%%%%%%%%%%%%%%%%%%%%%%%%%%%%%%%%%%%%%%%%%%%%%%%%%%%%%%%%%%%%%%%%%%%%%%%%%%%%%%%%%%
%%%%%%%%%%%%%%%%%%%%%%%%%%%%%%%%%%%%%%%%%%%%%%%%%%%%%%%%%%%%%%%%%%%%%%%%%%%%%%%%%%%%%%%%%%%%%%%%%%%%%%%%%%%%%%%%%%%%%%%%%%%%%
%%%%%%%%%%%%%%%%%%%%%%%%%%%%%%%%%%%%%%%%%%%%%%%%%%%%%%%%%%%%%%%%%%%%%%%%%%%%%%%%%%%%%%%%%%%%%%%%%%%%%%%%%%%%%%%%%%%%%%%%%%%%%

\subsection{Solutions over $\mathcal{M}_{\pm}$}

%%%%%%%%%%%%%%%%%%%%%%%%%%%%%%%%%%%%%%%%%%%%%%%%%%%%%%%%%%%%%%%%%%%%%%%%%%%%%%%%%%%%%%%%%%%%%%%%%%%%%%%%%%%%%%%%%%%%%%%%%%%%
%%%%%%%%%%%%%%%%%%%%%%%%%%%%%%%%%%%%%%%%%%%%%%%%%%%%%%%%%%%%%%%%%%%%%%%%%%%%%%%%%%%%%%%%%%%%%%%%%%%%%%%%%%%%%%%%%%%%%%%%%%%%%
%%%%%%%%%%%%%%%%%%%%%%%%%%%%%%%%%%%%%%%%%%%%%%%%%%%%%%%%%%%%%%%%%%%%%%%%%%%%%%%%%%%%%%%%%%%%%%%%%%%%%%%%%%%%%%%%%%%%%%%%%%%%%

The Euclidean sections solutions follow straightforwardly from the Lorentzian ones. For concreteness, consider the action \eqref{SE} and metric
on $\mathcal{M}_{+}$. Writing the metric as
\begin{equation}
ds^{2} = (1+r^{2})d\tau^{2} + \frac{dr^{2}}{1+r^{2}}+r^{2}d\varphi^{2}\,, \quad\quad\tau\in[0,\infty)
\label{Eglobales}
\end{equation}
and separating variables as
\be
\Phi_{+}(r,\tau,\varphi) \propto e^{+i \omega \tau + i l \varphi }\, h(\omega,l,r)\,,
\label{eans}
\ee
one readily finds
that
\begin{equation}
\label{h}
h(\omega,l,r)=f(-i\omega,l,r).
\end{equation}
Writing the general to solution to the KG equation as
\begin{equation}
\nn
\Phi_{+}(r,\tau,\varphi) =\sum_{l\in\mathbb{Z}} \int d\omega\, d_{\omega l} \, e^{i \omega \tau + i l \varphi}f(-i\omega,l,r)\,,
\end{equation}
and imposing
\be
\Phi_{+}(R,\tau,\varphi) =R^{\Delta-2} \phi_+(\tau,\varphi),
\label{ebc}
\ee
one finds
\begin{equation}
\label{Exp:InOut:ENNSol}
\Phi _{+}(r,\tau,\varphi)=\frac{R^{\Delta-2}}{4\pi^2 } \sum_{l\in\mathbb{Z}} \int d\omega d\tau' d\varphi' e^{i \omega (\tau-\tau') + i l (\varphi-\varphi')}  \phi_+(\tau',\varphi') \frac{f(-i\omega,l,r)}{f(-i\omega,l,R)}\,.
\end{equation}
In the Euclidean case the $\omega$-integral is well defined and the poles of the integrand lie on the imaginary axis as shown in figure  \ref{Exp:Fig:PolosB}.

As pointed out in \cite{SvRC},  considering the imaginary frequencies $\omega=\pm i\omega_{nl}^{_R}$ in \eqref{eans} yields, using \eqref{gnl} and \eqref{h},
$$\Phi_{+}(r,\tau,\varphi) \propto e^{\mp\omega_{nl}^{_R} \tau + i l \varphi }\, g_{nl}(r)\,.
$$
Since $\tau\in[0,\infty)$ for ${\cal M}_+$ one immediately notices that a linear combination of the $e^{-\omega_{nl}^{_R} \tau}$ modes could be added to  \eqref{Exp:InOut:ENNSol}\footnote{The converse happens for ${\cal M}_-$.}. Therefore, the solution to the KG equation over $\mathcal{M}_+$ satisfying \eqref{ebc} is
\begin{equation}
\label{Exp:InOut:ESol+}
\Phi_+(r,\tau,\varphi) = \frac{R^{\Delta-2}}{4\pi^2 } \sum_{l\in\mathbb{Z}} \int d\omega d\tau' d\varphi' e^{i \omega (\tau-\tau') + i l (\varphi-\varphi')}  \phi_+(\tau',\varphi') \frac{f(-i\omega,l,r)}{f(-i\omega,l,R)} + \sum_{\substack{n\in\mathbb{N} \\ l\in\mathbb{Z}}}E^{+}_{nl}\, e^{- \omega_{nl}^{_R} (\tau+ i T) + i l \varphi}  g_{nl}(r)\,.
\end{equation}
The solution over $\mathcal{M}_-$ can be found in a similar fashion and reads
\begin{equation}
\label{Exp:InOut:ESol-}
\Phi_-(r,\tau,\varphi) = \frac{R^{\Delta-2}}{4\pi^2 } \sum_{l\in\mathbb{Z}} \int d\omega d\tau' d\varphi' e^{i \omega (\tau-\tau') + i l (\varphi-\varphi')}  \phi_-(\tau',\varphi') \frac{f(-i\omega,l,r)}{f(-i\omega,l,R)} + \sum_{\substack{n\in\mathbb{N} \\ l\in\mathbb{Z}}} E^{-}_{nl}\, e^{\omega_{nl}^{_R} (\tau- i T) + i l \varphi}  g_{nl}(r)\,.
\end{equation}
The  $E^{\pm}_{nl}$ coefficients will be determined by the set of equations (\ref{Exp:InOut:CC}) (see \eqref{Coef}).
The existence of normalizable modes in Euclidean signature is a consequence of $\mathcal{M}_{\pm}$ being a half of the hyperbolic space,
containing either $\tau=\infty$ or $\tau=-\infty$, but not both. In the following we will consider non-trivial Euclidean boundary conditions vanishing
smoothly as $\tau\to\pm\infty$ and near $\Sigma^\pm$ as required by the SvR prescription \cite{SvRL}. The $e^{-i\omega_{nl}^{_R}   T}$ phases
in \eqref{Exp:InOut:ESol+}-\eqref{Exp:InOut:ESol-} are  convenient  for the contour $C$.

%%%%%%%%%%%%%%%%%%%%%%%%%%%%%%%%%%%%%%%%%%%%%%%%%%%%%%%%%%%%%%%%%%%%%%%%%%%%%%%%%%%%%%%%%%%%%%%%%%%%%%%%%%%%%%%%%%%%%%%%%%%%
%%%%%%%%%%%%%%%%%%%%%%%%%%%%%%%%%%%%%%%%%%%%%%%%%%%%%%%%%%%%%%%%%%%%%%%%%%%%%%%%%%%%%%%%%%%%%%%%%%%%%%%%%%%%%%%%%%%%%%%%%%%%%
%%%%%%%%%%%%%%%%%%%%%%%%%%%%%%%%%%%%%%%%%%%%%%%%%%%%%%%%%%%%%%%%%%%%%%%%%%%%%%%%%%%%%%%%%%%%%%%%%%%%%%%%%%%%%%%%%%%%%%%%%%%%%

\subsection{Matching the solutions}

%%%%%%%%%%%%%%%%%%%%%%%%%%%%%%%%%%%%%%%%%%%%%%%%%%%%%%%%%%%%%%%%%%%%%%%%%%%%%%%%%%%%%%%%%%%%%%%%%%%%%%%%%%%%%%%%%%%%%%%%%%%%
%%%%%%%%%%%%%%%%%%%%%%%%%%%%%%%%%%%%%%%%%%%%%%%%%%%%%%%%%%%%%%%%%%%%%%%%%%%%%%%%%%%%%%%%%%%%%%%%%%%%%%%%%%%%%%%%%%%%%%%%%%%%%
%%%%%%%%%%%%%%%%%%%%%%%%%%%%%%%%%%%%%%%%%%%%%%%%%%%%%%%%%%%%%%%%%%%%%%%%%%%%%%%%%%%%%%%%%%%%%%%%%%%%%%%%%%%%%%%%%%%%%%%%%%%%%

In order to perform the matching \eqref{Exp:InOut:CC} the asymptotic behavior of
\eqref{Exp:InOut:Sol}, \eqref{Exp:InOut:ESol+} and \eqref{Exp:InOut:ESol-} around $\Sigma^\pm$ is needed. These expressions can be easily
computed by residues since the sources vanish smoothly near the space-like boundaries \cite{SvRC,SvRL}. Closing the $\omega$-integrals appropriately
as shown in figure \ref{Exp:Fig:Polos} one finds
\begin{align}
\Phi_+(r,\tau,\varphi) &\sim \sum_{nl} \Big( E^{+}_{nl}\, e^{- \omega_{nl}^{_R} (\tau+ i T)} +R^{\Delta-2}  Res^R_{nl}\, \phi_{+;n(-l)}  e^{\omega_{nl}^{_R} (\tau+iT)}  \Big) e^{ i l \varphi}  g_{nl}(r),&\tau\sim0 \nn \\
\Phi_-(r,\tau,\varphi)  &\sim \sum_{nl} \Big( E^{-}_{nl}\, e^{\omega_{nl}^{_R} (\tau- i T)} + R^{\Delta-2} \,Res^R_{nl}\, \phi_{-;n(-l)} e^{- \omega_{nl}^{_R} (\tau-iT)}  \Big) e^{ i l \varphi}  g_{nl}(r),&\tau\sim0 \nn \\
\Phi_L(r,t,\varphi)&\sim  \sum_{nl} \Bigg( \Big( L_{nl}^{+} +i R^{\Delta-2} Res^R_{nl} \phi_{L;nl}^{*} \Big) e^{-i \omega_{nl}^{_R}  t} +   L_{nl}^{-} e^{i \omega_{nl}^{_R}  t} \Bigg) e^{ i l \varphi}  g_{nl}(r), &t\sim T  \nn \\
\Phi_L(r,t,\varphi) &\sim \sum_{nl} \Bigg( L_{nl}^{+} e^{-i \omega_{nl}^{_R}  t} +  \Big( L_{nl}^{-}+i R^{\Delta-2} Res^R_{nl}\,\phi_{L;n(-l)}  \Big) e^{i \omega_{nl}^{_R}  t} \Bigg) e^{ i l \varphi}  g_{nl}(r), &t\sim -T \nn
\end{align}
where in the first two lines
\begin{align}
\phi_{+;nl}&\equiv\frac{1}{2\pi}\int_0^\infty d\tau d\varphi \, e^{-\omega_{nl}^{_R} (\tau+ i T)+il\varphi}\, \phi_{+}(\tau,\varphi) \label{barphi+-} \, , &
\phi_{-;nl} & \equiv \frac{1}{2\pi}\int_{-\infty}^0 d\tau d\varphi\, e^{+\omega_{nl}^{_R} (\tau- i T)+il\varphi} \,\phi_{-}(\tau,\varphi)\,,
\end{align}
and for the Lorentzian piece
\be\nn
\phi_{L;nl} \equiv \frac{1}{2\pi}   \int_{-T}^{T} dt d\varphi \,e^{- i \omega_{nl}^{_R} t + i l \varphi}\, \phi_L(t,\varphi)  \,.
\ee
The residues $Res^R_{nl}$ are defined encircling the points $\omega=-\omega_{nl}^{_R}$ in a counterclockwise sense
\be\label{ResR}
Res^R_{nl} \equiv \frac{1}{2\pi i}\oint_{_{\omega=-\omega_{nl}^{_R}}} \frac{d\omega}{f(\omega,l,R)} \,.
\ee
From the orthogonality of $e^{ i l \varphi}$ and $g_{nl}(r)$,  equations (\ref{Exp:InOut:CC}) yield
\begin{align}
L_{nl}^{\pm} =& R^{\Delta-2} Res^R_{nl}\, \phi_{\mp;n(-l)}\,, \nn\\
E^{+}_{nl} =& R^{\Delta-2} Res^R_{nl}  \Big( i \phi^*_{L;nl} +  \phi_{-;n(-l)} \Big)\,, \nn \\
E^{-}_{nl} =& R^{\Delta-2} Res^R_{nl} \Big( i \phi_{L;n(-l)} +  \phi_{+;n(-l)} \Big)\,. \label{Coef}
\end{align}
These equations generalize the expressions found in \cite{SvRC} and reduce to them upon setting $\phi_\pm(\tau,\varphi)=0$. It is worth noticing that Euclidean boundary conditions turn
on normalizable modes in the Lorentzian section. We will show below that the $R^{\Delta-2}$ factors in \eqref{Coef} ensures a smooth $R\to\infty$ limit.

%%%%%%%%%%%%%%%%%%%%%%%%%%%%%%%%%%%%%%%%%%%%%%%%%%%%%%%%%%%%%%%%%%%%%%%%%%%%%%%%%%%%%%%%%%%%%%%%%%%%%%%%%%%%%%%%%%%%%%%%%%%%
%%%%%%%%%%%%%%%%%%%%%%%%%%%%%%%%%%%%%%%%%%%%%%%%%%%%%%%%%%%%%%%%%%%%%%%%%%%%%%%%%%%%%%%%%%%%%%%%%%%%%%%%%%%%%%%%%%%%%%%%%%%%%
%%%%%%%%%%%%%%%%%%%%%%%%%%%%%%%%%%%%%%%%%%%%%%%%%%%%%%%%%%%%%%%%%%%%%%%%%%%%%%%%%%%%%%%%%%%%%%%%%%%%%%%%%%%%%%%%%%%%%%%%%%%%%

\subsection{On shell action}

%%%%%%%%%%%%%%%%%%%%%%%%%%%%%%%%%%%%%%%%%%%%%%%%%%%%%%%%%%%%%%%%%%%%%%%%%%%%%%%%%%%%%%%%%%%%%%%%%%%%%%%%%%%%%%%%%%%%%%%%%%%%
%%%%%%%%%%%%%%%%%%%%%%%%%%%%%%%%%%%%%%%%%%%%%%%%%%%%%%%%%%%%%%%%%%%%%%%%%%%%%%%%%%%%%%%%%%%%%%%%%%%%%%%%%%%%%%%%%%%%%%%%%%%%%
%%%%%%%%%%%%%%%%%%%%%%%%%%%%%%%%%%%%%%%%%%%%%%%%%%%%%%%%%%%%%%%%%%%%%%%%%%%%%%%%%%%%%%%%%%%%%%%%%%%%%%%%%%%%%%%%%%%%%%%%%%%%%

In the present section the $R\to\infty$ limit is performed. The on shell evaluation of \eqref{ficmplx} results in a boundary term
that takes the form
\begin{align}
\nn
 S^0 &= -\frac{1}{2}\lim_{R\to\infty}\left[\ \int_{\partial_r\mathcal{M}} d{\sf t}\, d\varphi\, \sqrt{|\gamma_r|}\, \Phi \, n_r^{\mu} \partial_{\mu}\Phi\right]_{r=R}\\
 &=-\frac12\lim_{R\to\infty}\left[\int d\varphi\, (1+r^2)\, R^{\Delta-2} \left( -i \int_{-\infty}^0 d\tau \, \phi_{-} \, r \partial_{r} \Phi_{-} +  \int_{-T}^T dt \, \phi_L\, r \partial_{r}\Phi_{L}-i \int_{0}^{\infty} d\tau\, \phi_{+}\, r \partial_{r}\Phi_{+}\right)\right]_{r=R} ,
 \label{onchel}
\end{align}
where $\Phi(R,{\sf t},\varphi)=R^{\Delta-2}\,\phi({\sf t},\varphi)$ has been used. The three terms in \eqref{onchel} have identical radial behavior.
Taking the Lorentzian piece for concreteness, we now show that an expansion in $R\gg1$ picks contributions from both  $\phi_L$ and $L^\pm_{nl}$ terms in
\eqref{Exp:InOut:Sol}. The leading behavior of the radial derivative in \eqref{onchel} is
\begin{equation}
\label{expansion}
(1+r^2) \,R^{\Delta-2} \,\left( r \partial_{r}\Phi_{L} \right)\sim R^\Delta  \left( R^{\Delta-2}\phi_L  \frac{r\partial_{r}f(\omega,l,r)}{f(\omega,l,R)}
+  L_{nl}^{\pm}\, r \partial_{r}g_{nl}(r)  \right)\,,\qquad R\gg1\,.
\end{equation}
The contribution from the first term in \eqref{expansion} is the standard one. From the $r\gg1$ expansion of $f$
\begin{equation}
\label{f-expansion}
f(\omega,l,r) \sim A(\omega, l) r^{\Delta-2} + B(\omega, l) r^{-\Delta},
\end{equation}
where
\begin{equation}
\nn
A(\omega, l)\equiv \frac{\Gamma (\Delta-1 ) \Gamma(|l|+1)}{\Gamma \left(\frac{1}{2} (|l| +\Delta -\omega )\right) \Gamma \left(\frac{1}{2} (|l|+\Delta +\omega)\right)}\,, \qquad B(\omega, l)=\frac{\Gamma (1-\Delta ) \Gamma(|l|+1)}{\Gamma
 \left(\frac{1}{2} (|l|-\Delta -\omega +2)\right) \Gamma    \left(\frac{1}{2} (|l|-\Delta +\omega +2)\right)}\,,
\end{equation}
one finds, to leading order in $R$,
\begin{equation}
\label{expansionNN}
 R^{\Delta-2}\frac{r \partial_{r}f(\omega,l,r)}{f(\omega,l,R)}\Bigg|_{r=R}\sim \mathbb{S}[l,R] -2(\Delta-1)R^{-\Delta}\frac{B(\omega, l)}{A(\omega, l)}(1+ o(R^{-2}))\,.
\end{equation}
The first term $\mathbb{S}[l,R]$ being a regular series in $l$ is disregarded since, giving contact terms in configuration space, it can be subtracted by adequate counter-terms.

The $g_{nl}(r)$ expansion for $r\sim R\gg1$ is
\begin{equation}
\nn
g_{nl}(r) \sim A(\omega_{nl}^R,l)\, r^{\Delta-2} + B(\omega_{nl}^R,l)\, r^{-\Delta}\,,
\end{equation}
with $\omega_{nl}^R$ given by \eqref{modeR} so that $g_{nl}(R)=0$.
Thus, to leading order in $R$ one has $A(\omega_{nl}^R,l)\sim -B(\omega_{nl}^R,l) R^{-2\Delta+2}$ which inserted above gives
\begin{equation}\nn
g_{nl}(r) \sim -B(\omega_{nl}^R,l)\,R^{-\Delta} \left( \left(\frac{r}{R}\right)^{\Delta-2} - \left(\frac{r}{R}\right)^{-\Delta}\right)\,.
\end{equation}
With this expression at hand, on can readily see that to leading order in  $R$, the second term in \eqref{expansion} is
\begin{equation}
\label{Nexpansion}
 L_{nl}^{\pm}\, r\partial_{r}g_{nl}(r)\Big|_{r=R} \sim -2(\Delta-1)R^{-\Delta} \, B_{nl} \, Res_{nl}\phi_{\mp;n(-l)} (1+ o(R^{-2}))\,,
\end{equation}
where the $R$ dependence found in \eqref{Coef} was taken into account and using \eqref{ResR}
\begin{equation}\nn
B_{nl}\equiv \lim_{R\to\infty} B(\omega_{nl}^R, l)\,, \qquad Res_{nl}\equiv \lim_{R\to\infty}R^{\Delta-2 }Res^R_{nl} =\frac{1}{2\pi i} \oint_{\omega=-\omega_{nl}} d\omega \frac{1}{A(\omega,l)}.
\end{equation}
Inserting \eqref{expansionNN} and \eqref{Nexpansion} into \eqref{expansion} results schematically in a finite piece given by
\begin{equation}
\nn
\lim_{R\to\infty} (1+r^2) \,R^{\Delta-2} \, r \partial_{r}\Phi_{L} \sim  -2(\Delta-1)\left(\frac{B(\omega, l)}{A(\omega, l)}\phi_L + \, B_{nl} \, Res_{nl}\, \phi_{\pm}\right) \,.
\end{equation}
Carrying out similar calculations for the remaining pieces, the on-shell action becomes
\begin{align}
S^0 =&\, (\Delta-1)\Bigg(\sum_l\int_{{\cal F}} d\omega \, \phi_L(\omega,l)\phi^*_L(\omega,l) \frac{B(\omega,l)}{A(\omega,l)}\nn \\ &+2  \sum_{n l}\,\int dt d\varphi\, \phi_L(t,\varphi)\, \Big( \phi_{+;nl}\, e^{i\omega_{nl} t-il\varphi} + \phi_{-;n(-l)} \,e^{-i\omega_{nl} t+il\varphi}\Big)\, B_{nl} Res_{nl}\nn \\ & -i \sum_l \int d\omega \left( \phi_{+}(\omega,l) \phi^*_{+}(\omega,l) + \phi_{-}(\omega,l) \phi^*_{-}(\omega,l)  \right) \frac{B(-i\omega,l)}{A(-i\omega,l)} - 4\pi i \sum_{nl} \phi_{+;nl} \phi_{-;n(-l)} \,B_{nl} Res_{nl}\, \Bigg)\,,
\label{onshell}
\end{align}%(\Delta-1)
where
$$\phi_{+}(\omega,l)\equiv\frac{1}{2\pi} \int_0^\infty d\tau d\varphi\, e^{i\omega \tau + i l \varphi} \, \phi_{+}(\tau,\varphi)\, ,\qquad\phi_{-}(\omega,l)\equiv\frac{1}{2\pi} \int_{-\infty}^0 d\tau d\varphi\, e^{i\omega \tau + i l \varphi} \, \phi_{-}(\tau,\varphi)\,,$$
$$\phi_{L}(\omega,l)\equiv\frac{1}{2\pi} \int_{-T}^T dt d\varphi\, e^{-i\omega t + i l \varphi} \, \phi_{L}(\tau,\varphi)\,,$$
and
\begin{equation}
\label{BRes}
B_{nl} Res_{nl}=\frac{2\,(-1)^{n}\,\Gamma[1-\Delta]\,\Gamma[n+|l|+\Delta]}{n! \,(n+|l|)!\,\Gamma[1-n-\Delta]\,\Gamma[\Delta-1] }= 2(\Delta-1)\frac{\Gamma[\Delta+n+|l|]\Gamma[\Delta+n]}{n! (\Gamma[\Delta])^2\Gamma[n+|l|+1]}\,.
\end{equation}
Notice that \eqref{BRes} is positive for all $nl$, which can be readily seen from the rhs.

%Notice that $A(\pm\omega_{nl}, l)=0$, meaning that the lhs of \eqref{expansionNN} has poles in $\omega=\pm\omega_{nl}$.

%%%%%%%%%%%%%%%%%%%%%%%%%%%%%%%%%%%%%%%%%%%%%%%%%%%%%%%%%%%%%%%%%%%%%%%%%%%%%%%%%%%%%%%%%%%%%%%%%%%%%%%%%%%%%%%%%%%%%%%%%%%%
%%%%%%%%%%%%%%%%%%%%%%%%%%%%%%%%%%%%%%%%%%%%%%%%%%%%%%%%%%%%%%%%%%%%%%%%%%%%%%%%%%%%%%%%%%%%%%%%%%%%%%%%%%%%%%%%%%%%%%%%%%%%%
%%%%%%%%%%%%%%%%%%%%%%%%%%%%%%%%%%%%%%%%%%%%%%%%%%%%%%%%%%%%%%%%%%%%%%%%%%%%%%%%%%%%%%%%%%%%%%%%%%%%%%%%%%%%%%%%%%%%%%%%%%%%%

\subsection{Inner product and $n$-point correlation functions between excited states}
\label{Obs}

%%%%%%%%%%%%%%%%%%%%%%%%%%%%%%%%%%%%%%%%%%%%%%%%%%%%%%%%%%%%%%%%%%%%%%%%%%%%%%%%%%%%%%%%%%%%%%%%%%%%%%%%%%%%%%%%%%%%%%%%%%%%
%%%%%%%%%%%%%%%%%%%%%%%%%%%%%%%%%%%%%%%%%%%%%%%%%%%%%%%%%%%%%%%%%%%%%%%%%%%%%%%%%%%%%%%%%%%%%%%%%%%%%%%%%%%%%%%%%%%%%%%%%%%%%
%%%%%%%%%%%%%%%%%%%%%%%%%%%%%%%%%%%%%%%%%%%%%%%%%%%%%%%%%%%%%%%%%%%%%%%%%%%%%%%%%%%%%%%%%%%%%%%%%%%%%%%%%%%%%%%%%%%%%%%%%%%%%
%From \eqref{onshell} one can readily obtain every expectation value for ${\cal O}$ \textbf{EXPLICAR POR QUÉ $\Delta T=0$}

In the present section we devote to analyze the outcomes of \eqref{onshell}. We will evaluate the inner product between excited states and the  1,2-pt correlation functions.

~

\noindent {\sf Inner product}: As a consistency check, we will calculate the inner product from \eqref{onshell}, the result will match with \cite{GKP1,GKP2}.
The inner product between excited states can be computed by collapsing the Lorentzian piece, taking $\Delta T\to0$, in the absence of Lorentzian sources. This amounts to consider just
the third line in \eqref{onshell}\footnote{Notice that the $T$ dependence in \eqref{barphi+-} disappears in this limit.}. Explicitly one finds
\begin{align}
&\ln \langle \Psi^{\phi_+}| \Psi^{\phi_-} \rangle = \lim_{\Delta T\to0} i S^0 \nn\\
&= (\Delta-1)\left( \sum_l \int d\omega \left( \phi_{+}(\omega,l) \phi^*_{+}(\omega,l) + \phi_{-}(\omega,l) \phi_{-}^*(\omega,l)  \right) \frac{B(-i\omega,l)}{A(-i\omega,l)}  + 4\pi \sum_{nl} \phi_{+;nl} \phi_{-;n(-l)} B_{nl} Res_{nl} \right)\nn \\
&= (\Delta-1)\sum_l \int d\omega \Big( \phi_{+}(\omega,l) \phi^*_{+}(\omega,l) + \phi_{-}(\omega,l) \phi^*_{-}(\omega,l) + 2 \phi_{+}(\omega,l) \phi^*_{-}(\omega,l) \Big) \frac{B(-i\omega,l)}{A(-i\omega,l)} \nn \\
&= \frac 12 \int d\tau \, d\varphi \,d\tau' \,d\varphi' \, \Big(\phi_+(\tau,\varphi)+\phi_-(\tau,\varphi)\Big) {\cal P}(\tau,\tau',\varphi,\varphi') \Big(\phi_+(\tau',\varphi')+\phi_-(\tau',\varphi')\Big)\,,
\label{0pto}
\end{align}
where in the second line we have turned a sum over residues into an integral over $\omega$ and in the third line we have defined
\begin{align}\nn
{\cal P}(\tau,\tau',\varphi,\varphi')&\equiv \frac{\Delta-1}{2\pi^2} \sum_{l\in\mathbb{Z} }\int d\omega \, e^{i\omega(\tau-\tau')+i l (\varphi-\varphi')} \frac{B(-i\omega,l)}{A(-i\omega,l)}\nn \\
&= \frac{(\Delta-1)^2}{2^{\Delta-1}\pi}\Big(\cosh(\tau-\tau')-\cos(\varphi-\varphi')\Big)^{-\Delta}\quad,
\end{align}
which is the Euclidean 2-pt function on the cylinder, recovering GKPW \cite{GKP2}. Expression \eqref{0pto} is the explicit form of \eqref{botas} with ${\cal P}\equiv-G$ when Euclidean AdS
is parametrized as \eqref{Eglobales}.

~

\noindent {\sf 1-pt Correlation function}: The second line in \eqref{onshell} is the relevant one for computing the 1-pt function. The first derivative of \eqref{onshell}
with respect to the Lorentzian source in the $\phi_L\to0$ limits yields
\begin{align}
\frac{\langle \Psi^{\phi_+}| \mathcal{O}(t,\varphi)| \Psi^{\phi_-} \rangle}{\langle \Psi^{\phi_+}| \Psi^{\phi_-} \rangle}  &= - \frac{\delta S^0}{\delta \phi_{L}(t,\varphi)} \Bigg|_{\phi_{L}=0}\nn \\ &= -2 (\Delta-1) \sum_{n l}\, B_{nl} Res_{nl}\, \Big( \phi_{+;nl}\,\, e^{i\omega_{nl} t-il\varphi} + \phi_{-;n(-l)} \,\,e^{-i\omega_{nl} t+il\varphi}\Big)\,.
\label{1pto}
\end{align}
This non-zero result arises from considering non-zero boundary conditions $\phi_{\pm}$ in the Euclidean sections.

One could also notice that if we consider identical initial and final states in \eqref{1pto} ($\phi_+\equiv\phi_-^\star$) in the $T\to0$ limit, being ${\cal O}$ Hermitean, the result \eqref{1pto} should be real.
This condition yields
$$(\phi_{+;nl})^*=\phi_{-;n(-l)}\,,$$
or equivalently, from \eqref{barphi+-}, one finds
$$\phi_+(\tau,\varphi)=\phi_-(-\tau,\varphi)\, ,$$
which remarkably agrees with the Euclidean dual conjugate prescription given in \eqref{conjugate}.

~

\noindent {\sf Connected 2-pt function}:
Only the first line in \eqref{onshell} is relevant for the second derivative with respect to $\phi_L$. This gives the time ordered 2-point connected correlator, yielding
%A second derivative of the on-shell action keeps only the first line in \eqref{onshell} and 
\begin{align}
\frac{\langle \Psi^{\phi_+}|T[ \mathcal{O}(t,\varphi)\mathcal{O}(t',\varphi')]| \Psi^{\phi_-} \rangle}{\langle  \Psi^{\phi_+}| \Psi^{\phi_-} \rangle}\Bigg|_{c}  &\equiv -i  \frac{\delta^2 S^0 }{\delta \phi_{L}(t,\varphi)\delta \phi_{L}(t' ,\varphi')} \Bigg|_{\phi_{L}=0}\nn \\&= \frac{\Delta-1}{2\pi^2 i}\sum_{l\in \mathbb{Z}} \int_{{\cal F}} d\omega \,\,e^{-i\omega (t-t')+il(\varphi-\varphi')} \,\,\frac{B(\omega,l)}{A(\omega,l)} \nn \\ &=  \frac{(\Delta-1)^2}{2^{\Delta-1}\pi} \Big(\cos((t-t')(1-i\epsilon)) - \cos(\varphi - \varphi')\Big)^{-\Delta}\,.
\label{2pto}
\end{align}
which is independent of the boundary conditions $\phi_{\pm}$. This, perhaps unexpected result, will be explained in the next section.

% ~

% \noindent {\sf Connected n-pt function}: Further derivatives in \eqref{onshell} give trivial results since the action \eqref{ficmplx} is quadratic in the fields. 

%%%%%%%%%%%%%%%%%%%%%%%%%%%%%%%%%%%%%%%%%%%%%%%%%%%%%%%%%%%%%%%%%%%%%%%%%%%%%%%%%%%%%%%%%%%%%%%%%%%%%%%%%%%%%%%%%%%%%%%%%%%%
%%%%%%%%%%%%%%%%%%%%%%%%%%%%%%%%%%%%%%%%%%%%%%%%%%%%%%%%%%%%%%%%%%%%%%%%%%%%%%%%%%%%%%%%%%%%%%%%%%%%%%%%%%%%%%%%%%%%%%%%%%%%
%%%%%%%%%%%%%%%%%%%%%%%%%%%%%%%%%%%%%%%%%%%%%%%%%%%%%%%%%%%%%%%%%%%%%%%%%%%%%%%%%%%%%%%%%%%%%%%%%%%%%%%%%%%%%%%%%%%%%%%%%%%%
\subsection{Checking the coherent character of the initial/final states}
\label{check}
%%%%%%%%%%%%%%%%%%%%%%%%%%%%%%%%%%%%%%%%%%%%%%%%%%%%%%%%%%%%%%%%%%%%%%%%%%%%%%%%%%%%%%%%%%%%%%%%%%%%%%%%%%%%%%%%%%%%%%%%%%%%
%%%%%%%%%%%%%%%%%%%%%%%%%%%%%%%%%%%%%%%%%%%%%%%%%%%%%%%%%%%%%%%%%%%%%%%%%%%%%%%%%%%%%%%%%%%%%%%%%%%%%%%%%%%%%%%%%%%%%%%%%%%%
%%%%%%%%%%%%%%%%%%%%%%%%%%%%%%%%%%%%%%%%%%%%%%%%%%%%%%%%%%%%%%%%%%%%%%%%%%%%%%%%%%%%%%%%%%%%%%%%%%%%%%%%%%%%%%%%%%%%%%%%%%%%

In Sec. \ref{coher} we have shown that assuming \eqref{kaplan} and considering states of the form \eqref{estadoinicial-resultado}, the 1-pt function \eqref{kaplan2}
is linear in $\phi_\pm$. Moreover, an explicit expression for $\lambda^\pm_k$ can be written down using \eqref{eigenf-boundary} and \eqref{lambdak}.

On the other hand, using only the SvR prescription we computed the 1-pt function \eqref{1pto} that, compared to expression \eqref{kaplan2}, gives
\begin{equation}
\label{lambda} \lambda^\pm_{nl}\equiv-\sqrt{2\pi}\sqrt{ 2  (\Delta-1) B_{nl}  Res_{nl}}\,\phi_{\pm ;n(\mp l)}\,.
\end{equation}
It is now straightforward to check from \eqref{eigenf-boundary} and \eqref{BRes} that
$$N_{nl} = \sqrt{\frac{B_{nl}Res_{nl}}{2(\Delta -1)}}\,,$$
such that \eqref{1pto} remarkably coincide with the expression \eqref{kaplan2} up to a normalization factor which just rescales the operator.

In a similar fashion one could consider \eqref{estadoinicial-resultado} and \eqref{kaplan} and compute the inner product \eqref{0pto} and the
2-pt function \eqref{2pto} finding that they also match with our results.
The independence of \eqref{2pto} on $\phi_\pm$ can be better understood from the definition of the connected 2-point function
\begin{align}\nn
\frac{\langle \Psi^{\phi_+}|T[ \mathcal{O}(t,\varphi)\mathcal{O}(t',\varphi')]| \Psi^{\phi_-} \rangle}{\langle \Psi^{\phi_+}| \Psi^{\phi_-} \rangle}\Bigg|_{c}   \equiv& \frac{\langle \Psi^{\phi_+}| T[ \mathcal{O}(t,\varphi)\mathcal{O}(t',\varphi')]|\Psi^{\phi_-} \rangle}{\langle \Psi^{\phi_+}| \Psi^{\phi_-} \rangle} \\ & - \frac{\langle  \Psi^{\phi_+}| \mathcal{O}(t,\varphi)| \Psi^{\phi_-} \rangle}{\langle \Psi^{\phi_+}| \Psi^{\phi_-} \rangle}\frac{\langle  \Psi^{\phi_+}| \mathcal{O}(t',\varphi')| \Psi^{\phi_-} \rangle}{\langle \Psi^{\phi_+}| \Psi^{\phi_-} \rangle}\, .\nn
\end{align}
One can see that the $\phi_\pm$ dependence cancels out between the first and second terms in the rhs.

% A similar argument hold for connected $n$-point functions, where the expression consist of multiple terms that cancel such that for 
% $n>2$ every n-point function is vanishing. As we said before, from the gravity side this is manifest since the action is quadratic in the fields, so further derivatives are trivially zero.

Every result so far is consistent with the claim that the excited states \eqref{estadoinicial-resultado}, built by turning on Euclidean boundary conditions $\phi_\pm$ in the SvR prescription satisfy
\ba
\hat{a}_{nl} \,|\Psi^{\phi_-} \rangle &=& \Big( -\sqrt{2\pi}\sqrt{ 2  (\Delta-1) B_{nl}  Res_{nl}}\,\phi_{-;n(-l)} \Big)\,|\Psi^{\phi_-} \rangle\,,\\
\langle \Psi^{\phi_+} |\, \hat{a}^\dagger_{nl} &=& \langle \Psi^{\phi_+} | \, \Big( -\sqrt{2\pi}\sqrt{ 2  (\Delta-1) B_{nl}  Res_{nl}}\,\phi_{+;nl} \Big)\,,
\ea
which define $|\Psi^{\phi_\pm} \rangle$ as coherent states. Therefore, the states \eqref{estadoinicial-resultado} can be explicitly written in the bulk Hilbert (-Fock) space ${\cal H}_{AdS}$ as
\begin{equation}
\label{state-explicit}
|\Psi^{\phi_\pm} \rangle  = e^{\sum_{nl}\,2 \pi (\Delta-1)
 B_{nl}  Res_{nl}\phi_{\pm ;n(l)}\phi_{\pm ;n(-l)}} e^{ -\sum_{nl} \sqrt{ 4\pi  (\Delta-1)\,B_{nl}  Res_{nl}}\,\phi_{\pm ;n(\mp l)}\hat{a}_{nl}^\dagger}\,|0\rangle\,,
\end{equation}
in agreement with our expectations.

%%%%%%%%%%%%%%%%%%%%%%%%%%%%%%%%%%%%%%%%%%%%%%%%%%%%%%%%%%%%%%%%%%%%%%%%%%%%%%%%%%%%%%%%%%%%%%%%%%%%%%%%%%%%%%%%%%%%%%%%%%%%
%%%%%%%%%%%%%%%%%%%%%%%%%%%%%%%%%%%%%%%%%%%%%%%%%%%%%%%%%%%%%%%%%%%%%%%%%%%%%%%%%%%%%%%%%%%%%%%%%%%%%%%%%%%%%%%%%%%%%%%%%%%%
%%%%%%%%%%%%%%%%%%%%%%%%%%%%%%%%%%%%%%%%%%%%%%%%%%%%%%%%%%%%%%%%%%%%%%%%%%%%%%%%%%%%%%%%%%%%%%%%%%%%%%%%%%%%%%%%%%%%%%%%%%%%

\section{Concluding remarks}
\label{concluding}

%%%%%%%%%%%%%%%%%%%%%%%%%%%%%%%%%%%%%%%%%%%%%%%%%%%%%%%%%%%%%%%%%%%%%%%%%%%%%%%%%%%%%%%%%%%%%%%%%%%%%%%%%%%%%%%%%%%%%%%%%%%%
%%%%%%%%%%%%%%%%%%%%%%%%%%%%%%%%%%%%%%%%%%%%%%%%%%%%%%%%%%%%%%%%%%%%%%%%%%%%%%%%%%%%%%%%%%%%%%%%%%%%%%%%%%%%%%%%%%%%%%%%%%%%
%%%%%%%%%%%%%%%%%%%%%%%%%%%%%%%%%%%%%%%%%%%%%%%%%%%%%%%%%%%%%%%%%%%%%%%%%%%%%%%%%%%%%%%%%%%%%%%%%%%%%%%%%%%%%%%%%%%%%%%%%%%%

Through explicit holographic computations we have probed the Skenderis and van Rees proposal for \emph{excited states}.
We computed the time ordered 1- and 2-point functions in arbitrary states and have also found a noticeable expression to calculate
the inner product between them, which can be easily computed in the semi-classical approximation. The results support the hypothesis that the states built out by
turning on Euclidean boundary conditions are \emph{coherent} states. This could have also been argued by demanding consistency with other holographic recipes \cite{BDHM, BDHM2, kaplan}.

This scenario gives a more precise insight on the nature of the CFT-states that shall be associated to semiclassical gravity configurations
in the large $N$-limit, i.e, states where the spacetime geometry and the bulk fields behave classically. Our results are also in line with previous observations done
in the AdS/CFT literature \cite{bala1, marolf, sever}.

We have also learned that asymptotic boundary conditions play a crucial role when defining excited states in the Euclidean
path integral description of the wave functionals. The outcome of our computations is that the HH construction \cite{HH} can be generalized to
excited states in (quantum) gravity through the formula (\ref{wavef-g}), which properly generalized to include the gravitational degrees of
freedom reads
 \begin{equation}\nn
\Psi^{\phi_{-}, h_{-}}[\phi_{\Sigma^-}, h_{\Sigma^-}] \equiv \int^{(\phi_{\Sigma^-} \, ,\, h_{\Sigma^-})}_{(0\,,\, h_0)}
[\mathcal{D}\Phi]_{\phi_{-}} \,[\mathcal{D}(g, {\cal M}_-)]_{h_{-}} \, e^{-S^{}_-[\Phi, g] - S_{EH}[g]}\,.
\end{equation}%\label{wavef-g-HH}
The measure $[\mathcal{D} (g, {\cal M}_-)]_{h_{-}}$ stands for the $d+1$-dimensional spaces ${\cal M}_-$ that fit into the boundaries $\Sigma^- \cup {\cal S}^- $, endowed with Riemannian metrics $g$. They induce $d$-dimensional metrics  $h_{\Sigma^-}, h_{-}$ on the respective boundaries, as usual in Euclidean Quantum Gravity \cite{HH}. While $h_{\Sigma^-}$ can be arbitrary fixed, $h_{-}$ shall be defined as a suitable deformation of the conformal structure induced by Euclidean AdS on the asymptotic boundary $h_0 $, provided the deformation vanishes in the limit $\tau \to - \infty$.

In spacetimes with no asymptotic boundary, more investigation is required but
this observation suggest that this generalization could come up by inserting suitable extra boundaries in the Euclidean manifolds where one sums over. This idea will be explored in depth in a future work.

%%%%%%%%%%%%%%%%%%%%%%%%%%%%%%
\subsection*{Acknowledgements}
%%%%%%%%%%%%%%%%%%%%%%%%%%%%%%

GAS would like to thank B. van Rees for discussions and SAIFR-ICTP for
warm hospitality during the early stages of this work.
We acknowledge financial support from projects PICT 2012-0417, PIP0595/13 CONICET and X648 UNLP.

%%%%%%%%%%%%%%%%%%%%%%%%%%%%%%

\end{document}